\documentclass[superscriptaddress,12pt,tightenlines,aps,nofootinbib,pre]{revtex4-1}
\usepackage{color,epsfig,graphics,amssymb,amsmath,subeqnarray,graphicx,amsthm,subfigure,mathrsfs,bm,color,xcolor,soul}
\usepackage{natbib}

\def\createbib #1{\bibliographystyle{jsk}
\bibliography{#1}}

\usepackage{natbib}
\usepackage{graphicx}
\usepackage{amsmath}
\usepackage{amssymb}
\usepackage{natbib}
\usepackage{soul}
\usepackage{color}
\def\bea{\begin{equation}}
\def\eea{\end{equation}}

\usepackage{color}

\begin{document}
\title{The interaction of multiple bubbles in a Hele-Shaw channel}

\author{J. S. Keeler}
\affiliation{Mathematics Institute, University of Warwick,
Coventry, CV4 7AL, UK}
\email{jack.keeler@warwick.ac.uk}
\author{A. Gaillard}
\affiliation{Van der Waals-Zeeman Institute, University of Amsterdam, Science Park 904, Amsterdam, Netherlands}
\author{J. Lawless}
\affiliation{Department of Physics and Astronomy, University of Manchester, Manchester, M13 9PL}
\author{A. B. Thompson}
\affiliation{Department of Mathematics, University of Manchester, Manchester, M13 9PL,UK}
\author{A. Juel}
\affiliation{Department of Physics and Astronomy, University of Manchester, Manchester, M13 9PL}
\author{A. L. Hazel}
\affiliation{Department of Mathematics, University of Manchester, Manchester, M13 9PL,UK}

\begin{abstract}
  We study the dynamics of two air bubbles driven by the motion of a suspending viscous fluid in a Hele-Shaw channel with a small elevation along its centreline via physical experiment and numerical simulation of a depth-averaged model. For a single-bubble system we establish that, in general, bubble propagation speed monotonically increases with bubble volume so that two bubbles of different sizes, in the absence of any hydrodynamic interactions will either coalesce or separate in a finite time. However our experiments indicate that the bubbles interact and that an unstable two-bubble state is responsible for the eventual dynamical outcome: coalescence or separation. These results motivate us to develop an edge-tracking routine and calculate these weakly unstable two-bubble steady states from the governing equations. The steady states consist of pairs of `aligned' bubbles that appear on the same side of the centreline with the larger bubble leading. We also discover, through time-simulations and physical experiment, another class of two-bubble states which, surprisingly, consist of stable two-bubble steady states. In contrast to the `aligned' steady states, these bubbles appear either side of the centreline and are `offset' from each other. We calculate the bifurcation structures of both classes of steady states as the flow-rate and bubble volume ratio is varied. We find that they exhibit intriguing similarities to the single-bubble bifurcation structure, which has implications for the existence of $n$-bubble steady states.
\end{abstract}
\maketitle


\section{Introduction}
\label{sec:Introduction}

Identifying invariant objects (steady states, periodic orbits, invariant tori etc.) of high-dimensional, nonlinear systems and how they influence the transient dynamics is crucial in understanding how a system evolves towards an eventual dynamical outcome. One approach to identify these objects is to perform a number of initial-value problems (IVP), either experimentally or theoretically, and observe how the system behaves. The inherent disadvantage of this approach is that the outcome is binary; either the system settles to a stable invariant object, or long-term transient behaviour emerges. To capture unstable invariant objects, bespoke techniques are required, for example edge-tracking \citep{kerswell2014optimization} or parameter continuation \citep{kuznetsov2013elements,net2015continuation}. Although these invariant objects may be unstable, they still influence the dynamics of the system in a crucial way that would remain hidden in an IVP. In highly complex systems, such as the transition to turbulence in pipe flow \citep{kerswell2005recent,schneider2007turbulence,schneider2008laminar,kawahara2012significance}, these invariant solutions are often of high dimensionality and difficult to compute. We surmise, however, that these ideas are applicable to a large range of nonlinear systems and can be applied to systems which, although nonlinear and high-dimensional, are more amenable to theoretical and experimental analysis.

\begin{figure}
\centering
	\includegraphics[scale=0.8]{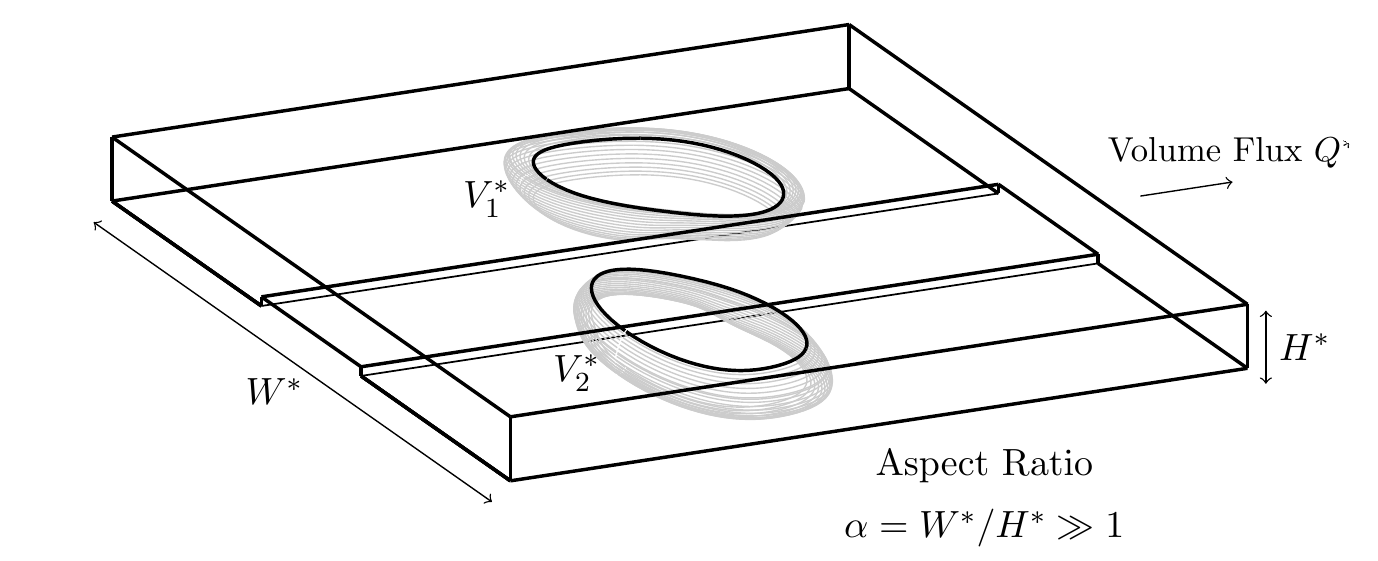}
	\caption{\small{A perturbed Hele-Shaw channel. Fluid is extracted a constant flux, $Q^*$, at one end, so that the bubbles propagate down the channel. The perturbation takes the form of a constant height and width rectangular rail at the bottom of the channel.}}        
	\label{fig:sketch_num_channel}
\end{figure}

\begin{figure}
	\centering
	\includegraphics[width=0.8\textwidth]{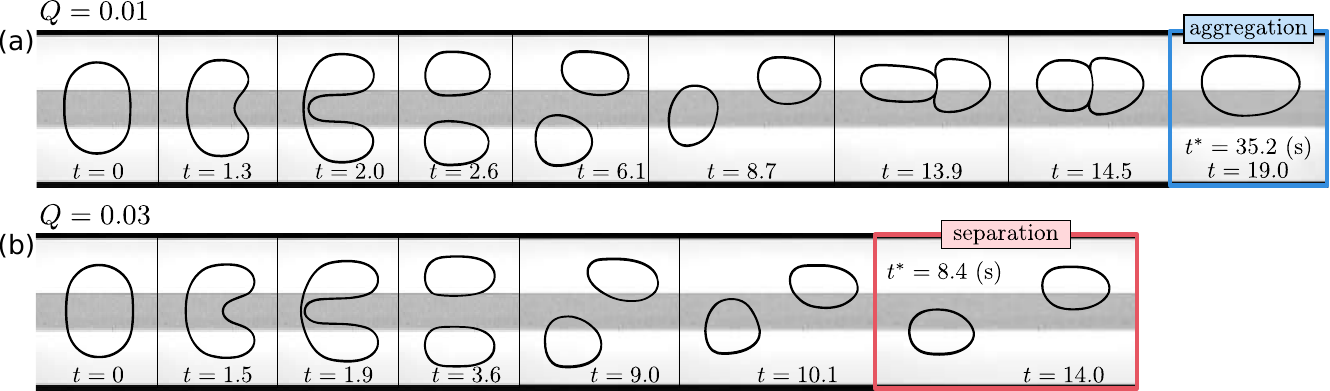}
	\caption{\small{Experimental time snapshots of the evolution of a bubble in a perturbed Hele-Shaw channel, as viewed from above. (a) $Q=0.02$. In this case, post-breakup, the smaller bubble crosses the rail and eventually coalesces with the larger bubble in an asymmetric configuration. (b) $Q=0.03$. In this case, post-breakup, the two bubbles separate indefinitely on either side of the rail.}}
	\label{fig:single_bubble_breakup}
\end{figure}

As a model `playground' to test these ideas we consider the steady state structure and transient dynamics of two finite air bubbles propagating in a Hele-Shaw channel with a prescribed depth perturbation when the surrounding fluid is extracted at a constant flow rate (see figure~\ref{fig:sketch_num_channel}). In a previous work \citep{gaillard2020life}, we showed that a single bubble may break up into two (or more) bubbles depending on its initial spatial configuration and on the flow rate and that, post breakup, the bubbles may either merge back into a single or compound bubble or separate indefinitely (see figure~\ref{fig:single_bubble_breakup}). A key result of this study was that the post breakup dynamics were strongly influenced by the existence of weakly unstable steady states that are specific to the two-bubble system. It was hence hypothesised that the complexity of the dynamics may increase with the number of bubbles, owing to the increase in the number of underlying (stable or unstable) steady states of the system.

A feature of this system is that the topology of the system changes when a bubble breaks up or two (or more) bubbles coalesce. Following such topological events, a different family of invariant solutions influence the transient dynamics. For a given system of, say, $n$-bubbles we might expect the steady states of the system to be related to the steady states of the lower-order $1,2,\cdots n-1$-bubble systems in such a way that a hierarchy of $1,2,\cdots n$-bubble states can be constructed from smaller bubble systems. The broad phenomenon of `lower-order' states interacting to form new coherent structures has been seen in other physical systems. For example, the interaction of solitons in water waves \citep[see, for example][]{drazin1989solitons} and nonlinear optics \citep[see, for example][]{akhmediev2005dissipative}, spatially localised states in convection systems \citep[see, for example][]{mercader2010covectons} and oscillons in granular particulate flow \cite[see, for example][]{umbanhowar1996localized}. A particular anomaly of our system is that we cannot smoothly move from a $n$-bubble state to a $m$-bubble state by continuation or branch-switching methods because the topologies of the systems are different. How the steady states of $n$ and $m$-bubbles relate to each other is therefore non-trivial and this system represents a rather different example of interacting localised states, from the previously highlighted. 

 The propagation of finite-air bubbles in a Hele-Shaw channel of uniform depth is a classical problem in fluid dynamics with a long and rich history. Transient behaviour and steady propagation modes have been investigated extensively in the case of a single-bubble using a mixture of analytical and numerical techniques \citep[see, for example][]{taylor1959note,tanveer1987surfacetension,lustri2018selection,tanveer1987stability,khalidunsteady,green2017effect} and experiments \citep[see, for example][]{kopf1988bubble,wang2014experimental,zhang2016particle,sirino2021experimental,maxworthybubble1986,madecthesis2021}. If a depth-perturbation is added to the bottom of the channel as shown in figure~\ref{fig:sketch_num_channel}, the range of existence and stability of steady propagation modes changes dramatically, as mapped out by \citet{franco2017bubble,franco2017propagation,keeler2019invariant,gaillard2020life}. The solution branches interact in a highly non-trivial manner, resulting in a number of different bifurcations and regions of bi-stability in the system; features absent when there is no geometric perturbation in the channel. Recently it has been shown that the transient behaviour of a single-bubble in a perturbed Hele-Shaw channel is heavily influenced by so-called `edge-states' of the system whose stable and unstable manifolds separate different dynamical outcomes \citep{keeler2019invariant,gaillard2020life}.

Although multi-bubble steady propagation modes in a Hele-Shaw channel have been studied in unperturbed channels, see, for example \cite{crowdy2012multiple,green2017effect,vasconcelos2015multiple,lustri2018selection}, these works have focused on steady state solution construction at zero surface tension and their significance to the underlying dynamics, including stability results, have not been investigated. Dipole models have been proposed to understand the dynamics of multiple bubbles in an infinite, unbounded Hele-Shaw cell, \citep{pumir1988,sarig2016dipole,green2018dipole}, by treating the bubbles as small and circular, and forming a system of ordinary differential equations describing the position of the individual bubbles based on interactions between each of them. The dipole model in an infinite, unbounded Hele-Shaw cell of uniform depth predicts that a single row of identical bubbles is neutrally stable but is prone to instability if `nudged' out of line. Also, relevant to this study, two rows of identical bubbles, located symmetrically about the horizontal centreline, is also neutrally stable, whilst two rows of bubbles which are located asymmetrically about the horizontal centreline is unstable, see \cite{pumir1988}. We remark that no stable multiple-bubble states have been observed in other confined systems and that in general the bubbles will always either separate or coalesce \citep{maxworthybubble1986,rohilla2020slug,madecthesis2021}. 

In this paper we concentrate on a two-bubble system in a depth-perturbed Hele-Shaw channel and investigate the existence of steady states and their dependence on the flow-rate and bubble volume. We calculate the two-bubble solution structure and find a number of two-bubble steady states, each playing a unique role in the underlying transient dynamics as the system parameters are varied. Surprisingly, we find that a stable steady state exists with a bubble on either side of the centreline and the smaller bubble leading. Furthermore, by comparing the two-bubble and single-bubble bifurcation diagrams, we uncover an underlying solution structure that may have implications for the existence of $n$-bubble steady states in general. We also make the observation that the dynamics of the two-bubble system are not necessarily dominated by the larger bubble, but rather it is the leading bubble that has the largest influence on the system.

The paper is organised as follows. In \S~\ref{sec:Methods_exp} and \S~\ref{sec:methods_num} we present the experimental and numerical methods used to investigate the dynamics of the system. In \S~\ref{sec:single_bubble_systems} we summarise the known results of the single-bubble system and extend these to explore the relationship between bubble speed and volume, which is fundamental to understanding the theoretical construction of two-bubble states. We then describe two classes of two-bubble states; {aligned} states where the bubbles have a similar vertical offset (\S~\ref{sec:aligned}), and {offset} states where the bubbles are staggered on either side of the rail (\S~\ref{sec:offset}). Next, in \S~\ref{sec:comparing_1_2_bubble_solutions}, we compare the solution structures of the two-bubble system to the single-bubble system before we conclude with a discussion of the implications of our results for $n$-bubble systems (\S~\ref{sec:discussion}).

\section{Experimental methods}
\label{sec:Methods_exp}

\begin{figure}
	\centering
	\vspace{0.5cm}
	\includegraphics[scale=1.0]{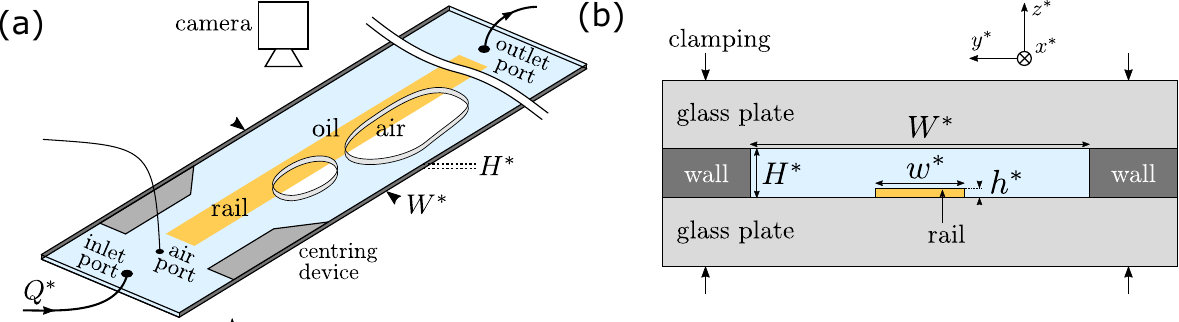}
	\caption{\small{(a) Schematic of the experimental setup and
            (b) experimental channel in cross-sectional view.}}
	\label{fig:exp_channel}
\end{figure}

We performed experiments in which two bubbles propagated through the channel from prescribed initial configurations imposed prior to flow initiation. The experimental Hele-Shaw channel presented in figure~\ref{fig:exp_channel} has been comprehensively described by \cite{gaillard2020life}. Thus, we only recall the salient details here. The channel consisted of two float glass plates separated by walls (strips of stainless steel shim), which were accurately positioned to make a channel of length $L^*=170$~cm, width $W^*=40 \pm 0.1$~mm and height  $H^* = 1.00 \pm 0.01$~mm, with an aspect ratio $\alpha = W^*/H^* = 40$. The channel was sealed with clamps and levelled horizontally to within 0.03$^{\circ}$. A centred rail of width $w^* = 10.0 \pm 0.1$~mm and thickness $h^* = 24 \pm 1$~$\mu$m consisted of a translucent adhesive tape strip bonded to the bottom glass plate, see figure~\ref{fig:exp_channel}(b). 

The channel was filled with silicone oil (Basildon Chemicals Ltd) of dynamic viscosity $\mu = 0.019$~Pa.s, density $\rho = 951$~kg/m$^3$ and surface tension $\sigma = 21$~mN/m at the laboratory temperature of $21\pm 1$~$^{\circ}$C, and connected to oil reservoirs through inlet and outlet ports located at the upstream and downstream ends of the channel, respectively (see figure~\ref{fig:exp_channel}(a)). Flow in the channel was imposed by injecting oil through the inlet port with constant volume flux $Q^*$ using a bank of three syringe pumps, and letting oil escape through the outlet port. Air bubbles were generated by injecting prescribed volumes of air in the channel through an air port positioned slightly downstream of the inlet port; see \citet{gaillard2020life} for details on the bubble generation protocol. Once formed, the bubbles were propagated through a centring device consisting of a section of channel of reduced width followed by a region of linear expansion, as shown schematically in figure~\ref{fig:exp_channel}(a).

Experiments were performed with pairs of bubbles, each of prescribed
area as measured from above, which were arranged in reproducible initial configurations in terms of their shapes and relative positions. We distinguish `aligned’ initial bubble configurations from `offset' configurations. The former correspond to axially aligned bubbles with both bubbles either positioned symmetrically about the channel centreline (`on-rail') or asymmetrically (`off-rail') but on the same side of the rail (figure~\ref{fig:exp_channel}(a)). In the `offset' configuration, two off-rail bubbles are positioned on opposite sides of the rail as shown schematically in figure~\ref{fig:computational_domain}. These initial bubble configurations were generated using two different experimental protocols described in appendix~\ref{sec:app_exp_protocols}.

Bubbles were propagated from their initial configuration at a constant dimensionless flow rate $Q = \mu U_0^* / \sigma$ where $U_0^* = Q^*/(W^* H^*)$ is the average oil velocity in an equivalent channel without the rail. The bubbles were filmed in top-view using a CMOS camera mounted on a motorised stage, which translated at a constant velocity value chosen to ensure that the bubbles remained within the field of view of the camera for the duration of the experiment. We refer to each initial bubble with a numerical index in order of decreasing size, so $i=1$ corresponds to the largest bubble. The projected area $A_i^*$ ($i=1,2$) and centroid position of each bubble were measured from the bubble contour detected using an edge detection algorithm. The distance between the two bubbles is quantified by $D = 2D^*/W^*$ where $D^*$ is the dimensional distance between the centroids of the two bubbles. Unless otherwise specified, the combined bubble size is $A_{\mathrm{total}} = A_1 + A_2 = 0.54^2 \pi$ which is the size of the single bubbles used in \citet{gaillard2020life}, where $A_i = A_i^* / (W^*/2)^2$ ($i=1,2$) is the non-dimensional area of each bubble. We investigated initial configurations with either the larger or smaller bubble in the lead position and report results in terms of the initial fractional size of the lead bubble given by the ratio $A_r = A_{\mathrm{lead}}/A_{\mathrm{total}}$, where $A_{\mathrm{lead}}$ may equal $A_1$ or $A_2$ depending on the initial order of the bubbles. For convenience, we sometimes also refer to the ratio of the bubble areas in the form $A_1:A_2$. For example, bubbles of volume ratio 2:1 have $A_r = 2/3$ if the larger bubble is (initially) leading and $A_r = 1/3$ otherwise. 

 In the experiment, we only measure the projected area directly during bubble propagation. For a fixed volume of injected air, the projected area of the bubble can decrease sharply when flow is initiated because of air compression which increases with flow rate as the associated pressure head increases.  However, the bubble retains an approximately constant projected area during each experiment, with a small increase of less than 7\% at the highest flow rates. Conversely, the presence of lubricating oil films separating the bubble from the top and bottom plates, whose thickness increases with increasing flow rate, tends to increase the projected area of the bubble. These effects are discussed in detail in \citet{gaillard2020life} and a suitable calibration of the injected volume of air was performed to obtain propagating bubbles with the required prescribed areas $A_i$ ($i=1,2$). 

\section{Mathematical model and numerical methods}
\label{sec:methods_num}

\begin{figure}
	\centering
	\includegraphics[scale=0.35]{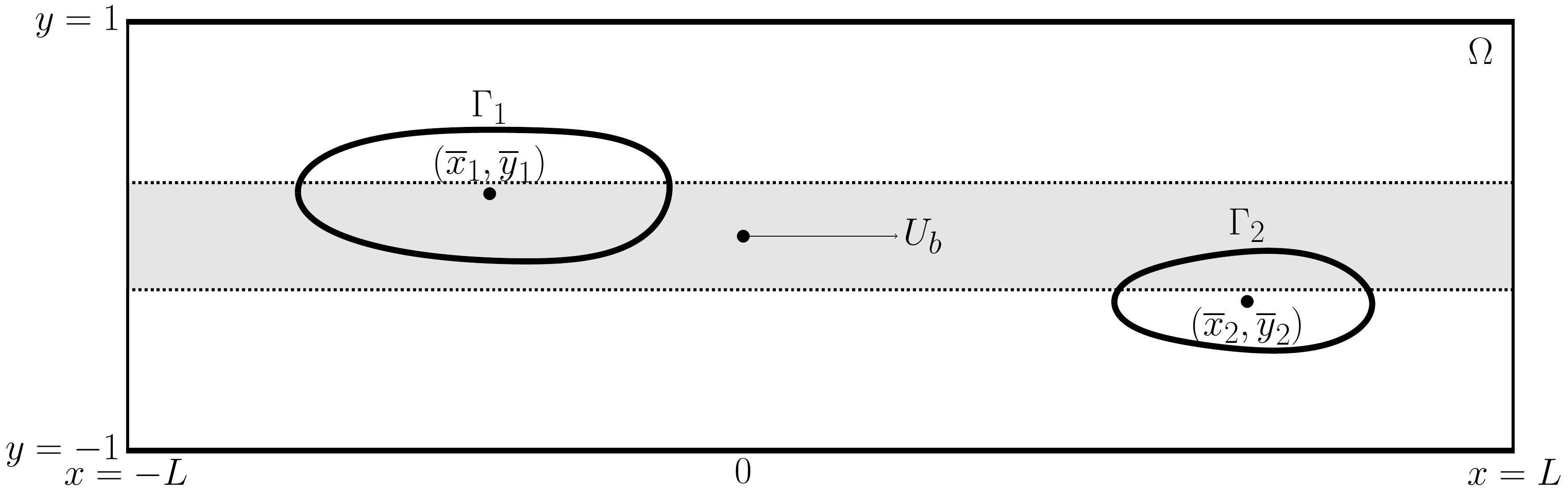}
	\caption{\small{Sketch of the nondimensional computational domain, which is in a frame of reference centred on the overall centre of mass that moves with speed $U_{\mathrm{b}}$. The horizontal domain is truncated at a value of $x = L$ (typically $L=6$ in our calculations) and the boundaries of the rectangular rail are marked using dotted horizontal lines. The fluid domain is denoted $\Omega$, and the two air bubble boundaries are denoted $\Gamma_1$ and $\Gamma_2$. The centroids of the bubbles are denoted $(\overline{x}_i,\overline{y}_i)$.}}
	\label{fig:computational_domain}
\end{figure}

The depth-averaged model for the propagation of multiple bubbles in our Hele-Shaw channel has been previously described and we only summarise its key features below.  Our approach extends that of \cite{mccleantension} to account for a non-uniform channel height and has been used extensively in studies of the propagation of a semi-infinite air finger \citep{thompson2014multiple,franco2016sensitivity}, single closed air bubbles \citep{franco2017bubble,franco2017propagation,keeler2019invariant} and most recently single and multiple air bubbles \citep{gaillard2020life}. We use the model to compute steady states of the system, calculate their linear stability and perform numerical time-simulations. 

We work in a frame moving with the centroid position of the entire collection of bubbles and non-dimensionalise the physical system shown in figure \ref{fig:exp_channel} using $W^*/2$ and $H^*$ as characteristic length scales in the $(x^*,y^*)$ plane and $z^*$ direction, respectively, and $U_0^* = Q^*/(W^* H^*)$ as the velocity scale. The resulting nondimensional computation domain is shown in figure~\ref{fig:computational_domain}.

 The two-dimensional depth averaged lubrication model reduces to an equation for the pressure in the fluid domain 
\begin{equation}
\nabla\cdot (b^3 \nabla p) = 0\qquad (x,y)\in \Omega,
\label{bulk}
\end{equation}
where the mobility $b(y)$ represents the variable depth of the channel, modelled as a smoothed tanh profile
\begin{equation}
b(y) = 1 - \frac{1}{2}h\left[\tanh(s(y + w)) - \tanh(s(y - w))\right],
\end{equation}
	where $h = h^*/H^*$ and $w = w^*/W^*$ are the non-dimensional height and width of the rail respectively; and $s$ sets the `sharpness' of the sides of the rail, as shown in figure \ref{fig:computational_domain}. We use $h=0.024$ and $w=0.25$ consistent with experiments and we choose $s=40$ \citep{thompson2014multiple}. We impose no-penetration conditions on the upper and lower walls, which yield $p_y = 0$ on $y=\pm 1$. The pressure is fixed to zero at the inflow, and a non-zero constant at the outflow to ensure the dimensionless volume flux is consistent with the inflow dimensionless volume flux. 

Equations are solved in the reference frame moving at velocity $\textbf{U}$ and we assume that the air bubble fills the height of the channel so that the kinematic boundary conditions on the contour of each bubble denoted by $\textbf{R}_i$ (where $i=1,2,\ldots$ indicates the $i$\textsuperscript{th} bubble in decreasing size order) is given by
\begin{equation}
\frac{\partial\,\textbf{R}_{i}}{\partial\,t}\cdot \textbf{n}_i + \textbf{U}\cdot\textbf{n}_i + b^2\nabla p\cdot \textbf{n}_i = 0,
\label{kinematic}
\end{equation}
where $\textbf{n}_i$ is the unit normal vector directed away from the $i$\textsuperscript{th} bubble and $\textbf{U} = (U_{\mathrm{b}},0)$ is the velocity of the centre of mass of the system along $x$. The centre of mass speed, $U_{\mathrm{b}}$, is an unknown in the problem which is obtained by requiring that the $x$ coordinate of the centre of mass of the system remains at zero. The dynamic boundary condition on each bubble is 
\begin{equation}
p_i - p = \frac{1}{3\alpha Q}\left(\frac{\kappa}{\alpha} + \frac{1}{b(y)}\right),
\label{eq:dynamic}
\end{equation}
where $\kappa$ denotes the curvature of the bubble in the $(x,y)$ plane and the effects of the variable depth on the transverse curvature are accounted for by the $1/b(y)$ term. The pressure $p_i$ in each bubble is not known \textit{a priori} and is determined by ensuring that the dimensionless bubble volume $V_i$ remains constant, where the volume $V_i$ is defined via
\begin{equation}
V_i = \int_{\Gamma_i} b(y) \, \mathrm{d}x \, \mathrm{d}y=  \int_{\partial \Gamma_i} x b(y) \, \mathrm{d}y, \quad A_i= \int_{\Gamma_i} \, \mathrm{d}x \, \mathrm{d}y.
\label{eq:Vi}
\end{equation}
where $\Gamma_i$ is the interior of bubble $i$ and $\partial \Gamma_i$ its bounding curve. The total bubble volume $V_{\mathrm{total}}=V_1+V_2$ is set to $0.54^2\pi$ unless specified otherwise, consistent with the experiments of \cite{gaillard2020life}. In the model, the dimensionless volume and area of a given bubble are almost identical because we assume that the bubble fills the entire channel height, which differs from 1 by at most 2.5 \%. In the experiment, we measure the size of the experimental bubbles by their projected areas $A_i$ ($i=1,2$) because the air volume required to yield a bubble of fixed projected area varies with flow rate as discussed in \S \ref{sec:Methods_exp}. 

We solve the system of equations, \eqref{bulk}-\eqref{eq:Vi} on the domain shown in figure \ref{fig:computational_domain} to determine $p$, $\textbf{R}_i$, $p_i$ and $U_{\mathrm{b}}$ and we use the flow rate $Q$ and bubble volumes $V_i$ as control parameters. The spatial discretisation of the equations is obtained by using a finite-element method, see appendix~\S\ref{sec:app_finite_elements}. When performing time-simulations, we use a procedure detailed in appendix~\S\ref{sec:app_topology_changes} to account for the topology changes that may occur, such as bubble breakup and coalescence events. Stable and unstable steady solutions of the governing equations are calculated using Newton's method. Convergence of this method requires a good initial guess for the bubble configuration. For {stable} steady states, an initial guess can be obtained by performing a time-simulation from an initial condition where the system converges towards the stable state. For {unstable} steady states however, finding a good initial guess requires bespoke methods for each individual state, see appendix~\ref{sec:app_interval_bisection}. Once a stable or unstable steady state has been identified for a given set of control parameters, we use continuation methods to map the solution space as the control parameters are varied.

\subsection{Two-bubble metrics}
\label{sec:control_parameters}

We characterise the two-bubble system by the coordinates ($\overline{x}_i$, $\overline{y}_i$) of the centroids of each bubble given by
\begin{equation}
  (\overline{x}_i,\overline{y}_i) = \frac{1}{A_i} \left( \int_{\Gamma_i} x \, \mathrm{d}x \, \mathrm{d}y , \int_{\Gamma_i} y \, \mathrm{d}x \, \mathrm{d}y \right),
\label{eq:centroid}
\end{equation}
where $A_{i}$ is defined in equation (\ref{eq:Vi}).
From these, we compute the distance $D$ between the centroids of each bubble, given by
\begin{equation}
D = \sqrt{(\overline{x}_1 - \overline{x}_2)^2 + (\overline{y}_1 - \overline{y}_2)^2},
\label{eq:D}
\end{equation}
as well as the $y$-coordinate of the centre of mass of the system defined by
\begin{equation}
  \overline{Y} = \frac{V_1\overline{y}_1 + V_2\overline{y}_2}{V_1 + V_2},
\label{eq:Y}
\end{equation}
which we refer to as the offset of the two-bubble system. In time-simulations, these quantities will evolve as functions of time, $\overline{x}_i(t)$, $\overline{y}_i(t)$, $D(t)$ and $\overline{Y}(t)$, and will be used to characterise the state of the system.

In numerical time-simulations, the initial shape of each bubble was chosen to be an ellipse with contour coordinates
\begin{equation}
\textbf{R}_i(t=0) = (\overline{x}_{i}(0) + \ell_i\cos\theta , \,  \overline{y}_{i}(0) + d_i\sin\theta),\quad 0\leq\theta<2\pi.
\end{equation}
In all the numerical time-simulations presented in this paper, the volume ratio was $V_1 : V_2 = 2:1$ and we chose initially slender bubbles with $d_1 = 0.3$ and $d_2 = 0.2$ so that the bubbles did not break up before they interacted. The values of $\ell_i = V_i/\pi d_i$ were set to ensure the prescribed volumes. 

\section{Results}

\subsection{Single-Bubble Systems}
\label{sec:single_bubble_systems}

\begin{figure}
	\centering
	\includegraphics[scale=0.35]{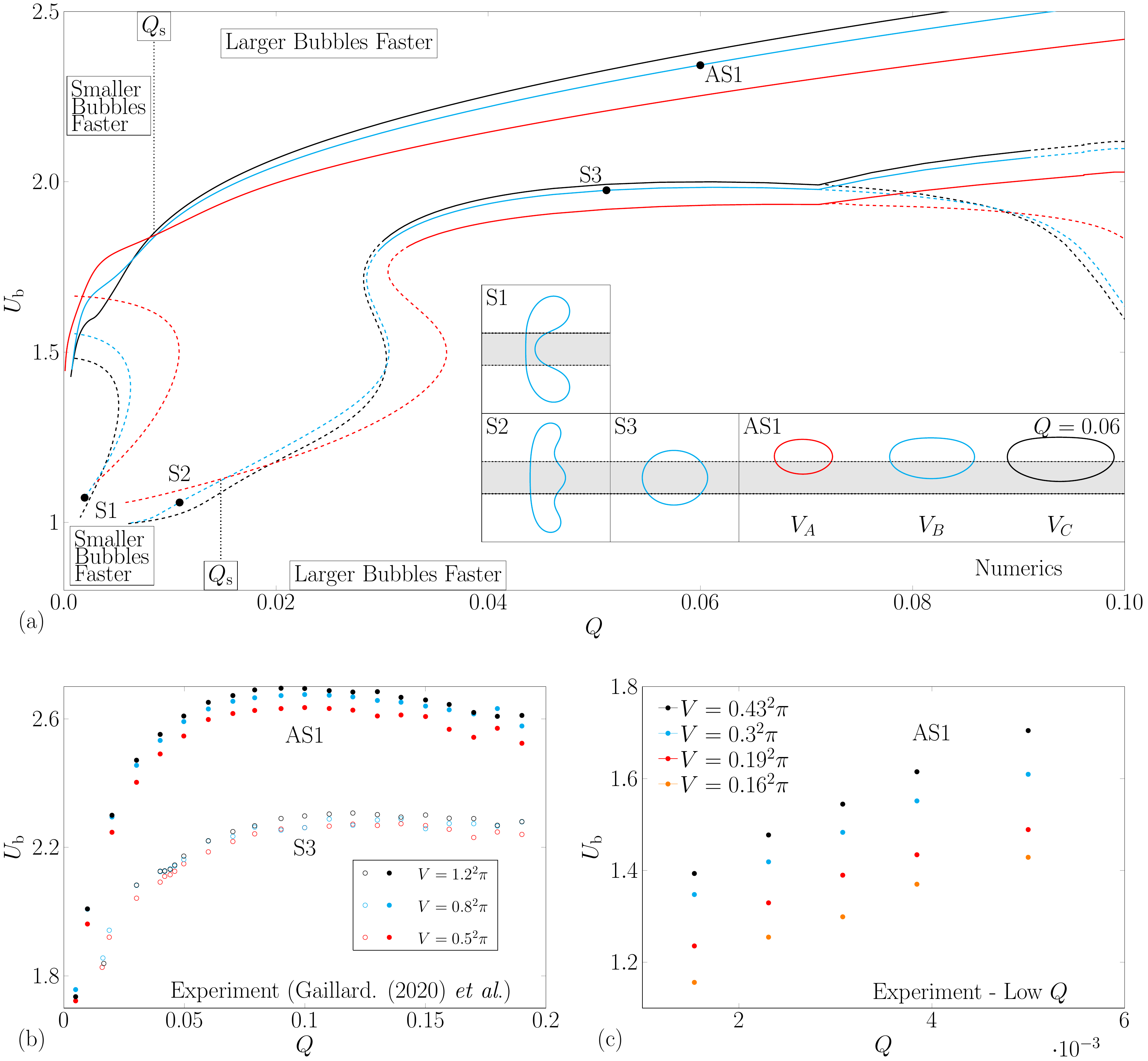}
	\caption{\small{Single-bubble solution space. The velocity $U_{b}$ is plotted as a function of dimensionless flow rate $Q$ for (a) the different solution branches of the theoretical model and (b,c) for the stable steady modes of propagation observed in experiments, where (c) focuses on the lowest flow rates. The experimental data presented in (b) is reproduced from \cite{gaillard2020life} with permission. The inset panels in (a) correspond to bubble profiles specified by solid circular markers on the branches. The bubbles have volumes $V=\pi 0.54^2,\pi 0.46^2,\pi 0.35^2$, indicated by the different colours of the branches. In the theoretical results, the flow rate $Q_{\mathrm{s}}$ indicates where a `switchover' occurs in the relative speeds of larger and smaller bubbles. There is no evidence for the existence of $Q_{\mathrm{s}}$ in the experiments. The hollow circular markers in (b) correspond to the symmetric S3 state and the solid circular markers in (b) and (c) correspond to the asymmetric AS1 state.}}
	\label{fig:Qswitch}
\end{figure}

Before we discuss two-bubble systems, we present an overview of the steady propagation of single bubbles and examine the influence of bubble volume over our range of interest. Figure \ref{fig:Qswitch} presents theoretical, panel (a), and experimental results, panels (b,c), for the dimensionless speeds, $U_{b}$, of individual bubbles as functions of dimensionless flow rate $Q$. The different colours correspond to different bubble volumes and we find that the structure of the steadily propagating solutions in the theoretical model is independent of bubble volume within the range investigated. We label the different solution branches as in \cite{keeler2019invariant} and \cite{gaillard2020life}. For our region of interest, there are three distinct solutions: a stable asymmetric bubble, AS1, that exists for all flow-rates; an unstable symmetric double-tipped bubble, S1, that exists for small flow-rates; and an alternative symmetric bubble, S2/S3, that exists for larger flow-rates and can be stable or unstable. Inset snapshots in figure \ref{fig:Qswitch}(a), show the shapes of the bubble at flow rates indicated by dots on the solution branches. For an intermediate range of flow rates, the system is bistable with both symmetric and asymmetric, stable propagation modes available.

The experimental data shown in figure \ref{fig:Qswitch}(b) correspond to the symmetric and asymmetric states in the bistable regime and we find the following general trends in both experiments and the theoretical model:
(i) the dimensionless speed $U_{\mathrm{b}} = U^{*}_{b}/U^{*}_{0}$, of the bubble relative to the average speed of the surrounding fluid, typically increases with flow rate (note that for high flow rates the relative bubble speed saturates for the symmetric state and decreases slightly for the asymmetric state); (ii) for a fixed flow rate, the dimensionless speed increases with bubble volume, for the range of volumes investigated here. At low flow rates in the theoretical model, however, the opposite trend is observed.
 
 The result that larger bubbles travel faster at a fixed flow rate follows from the theoretical analysis of \cite{taylor1959note}, under the assumption of fixed bubble width. The same theory predicts that for a fixed volume, wider bubbles travel faster. The explanation for these results is that either increasing bubble width for fixed volume, or increasing volume for fixed width leads to increased viscous dissipation. Increased viscous dissipation is balanced by an increase in the work done by fluid pressure on the bubble, which results in a higher local fluid pressure gradient. The increased pressure gradient leads to a higher local fluid velocity around the bubble, which leads to faster bubble speeds via the kinematic condition, equation (\ref{kinematic}). Related results have also been found for buoyant rise of bubbles in Hele-Shaw cells \citep{maxworthybubble1986} in which the bubble speed also increases with bubble width. In that case, however, the increase in viscous dissipation is balanced by an increase in the work done by the buoyancy force, which itself increases with bubble volume.
 Similar arguments can explain why in the bistable regime the asymmetric bubble travels faster than the symmetric one, for a fixed flow rate. The symmetric bubble spans the rail which means that it displaces a smaller area of fluid within each cross-section as it propagates, leading to lower dissipation, lower local pressure gradient and hence a lower bubble speed.
 
 At lower flow rates ($Q < 0.02$), there is a qualitative disagreement between the model and the experimental results. In the experiments, the relative propagation speed of single bubbles is an increasing function of the bubble size at {all} values of the flow rate investigated, see figure \ref{fig:Qswitch}(c). In the model, for both the asymmetric and symmetric solutions there is a critical flow rate below which the relative bubble speed decreases as the bubble volume increases. The value of $Q$ where this trend `switches' over is denoted $Q_{\mathrm{s}}$. This discrepancy has been previously noted by \cite{maxworthybubble1986} who states ``It is also clear also that the theory overestimates the bubble velocities for the smaller widths, having the wrong behaviour as $D$ (the diameter) $\to 0$'' \cite[pp.108]{maxworthybubble1986}. 

We accept, therefore, that for small bubbles and low-rates the model does not reflect the experiments and we confine the majority of our analysis to $Q>Q_{\mathrm{s}}$ where the model and experiment predict the same speed-volume relationship.

\subsection{Two-Bubble Systems}
\label{sec:two_bubble_systems}

 From the results for single bubbles, \S
 \ref{sec:single_bubble_systems}, we know that two bubbles of
 different volumes will always travel at different speeds when the
 surrounding fluid moves at a fixed flow rate $Q$. Hence, in the
 absence of any hydrodynamic interactions and irrespective of the
 initial separation, two different bubbles will either coalescence in
 finite time or separate indefinitely depending on whether the slower
 bubble is initially ahead or behind the faster, respectively. In this
 section, we demonstrate that bubbles interact hydrodynamically,
 which results in the existence of stable and unstable two-bubble
 steady modes of propagation. As far as we are aware, such modes have
 not been observed for bubble of different sizes in related confined
 systems. In those cases, the bubbles will either separate or coalesce
 \citep{maxworthybubble1986,madecthesis2021}.

 We first consider aligned states, see \S \ref{sec:aligned}, in which the two bubbles have similar  centroid $y-$coordinates ($\overline{y}_1 \approx \overline{y}_2$), and then offset states, see \S \ref{sec:offset}, where the two bubbles are on opposite sides of the rail ($\overline{y}_1 \overline{y}_2 < 0$). 

 \subsubsection{Aligned Bubbles}
\label{sec:aligned}

 Bubble pairs propagating from aligned initial configurations were prepared experimentally using the protocol outlined in appendix~\ref{sec:app_exp_protocol_aligned}. For both symmetric (`on rail') and asymmetric ('off rail') pairs of bubbles, when the larger bubble was initially placed behind the smaller one, the bubbles always coalesced. Hence, any hydrodynamic interactions were not sufficient to prevent the behaviour predicted from the single-bubble results, in which larger bubbles move faster.

\begin{figure}
	\centering
	\includegraphics[scale=1.0]{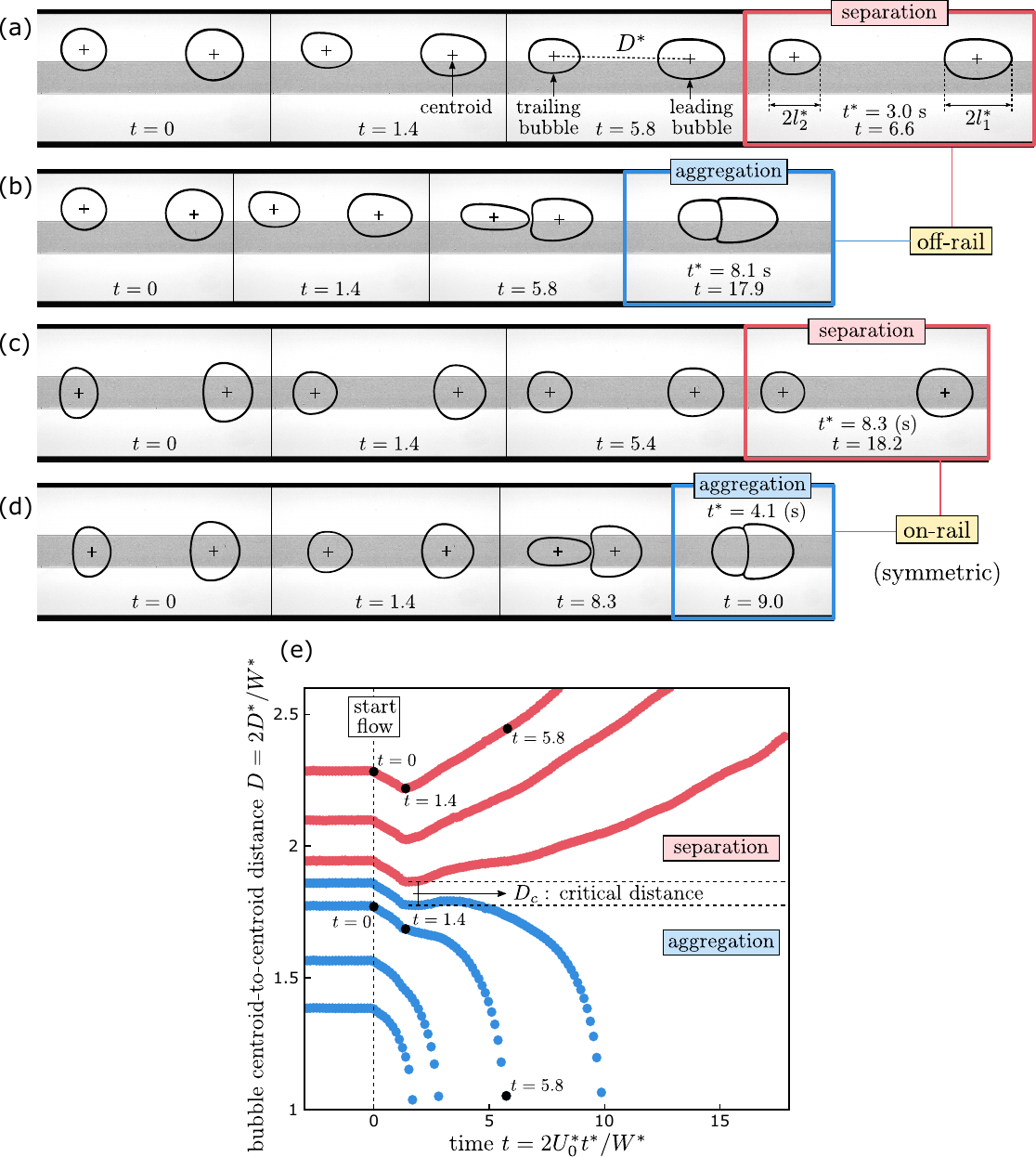}
	\caption{\small{Time sequences of two-bubble evolution for different initial bubble-pair configurations, where the leading bubble is larger than the trailing bubble: (a, b) aligned asymmetric bubbles overlapping the rail from one side, (c, d) aligned bubbles straddling the rail symmetrically about its centreline. The non-dimensional initial distance between the centroids of each bubble (which are indicated by crosses) is $D(t=0)=2.10$ (a), $1.77$ (b), $2.39$ (c) and $1.98$ (d). The flow rate is $Q = 0.04$ ($Q^* = 106$~ml/min), the total bubble area is $A_1 + A_2 = 0.54^2 \pi$ and the bubble size ratio is $A_1/(A_1+A_2) = 0.60$. Each row of top view images shows the evolution of the system in terms of the non-dimensional time $t = 2 U_0^* t^*/ W^*$ elapsed since flow initiation at $t=0$. The dimensional time $t^*$ is indicated in the last snapshot of each time-sequence. (e) Time evolution of the non-dimensional distance $D = 2D^*/W^*$ between the centroids of two asymmetric bubbles propagated from different initial separation distances. The two curves with time labels correspond to the time-sequences shown in (a,b). $D_{\mathrm{c}}$ is the critical bubble distance delineating aggregation (blue curves) and separation (red curves).}}
	\label{fig:exp_aligned}
\end{figure}

Figure~\ref{fig:exp_aligned} shows propagation experiments in which
the larger bubble is initially leading. For a sufficiently large
initial separation, the bubbles separate indefinitely (figures
\ref{fig:exp_aligned}(a), asymmetric, and \ref{fig:exp_aligned}(c),
symmetric), as predicted from the single-bubble results. For smaller
initial separation distances, however, the two bubbles aggregate to
form a compound bubble (figures \ref{fig:exp_aligned}(b), asymmetric,
and \ref{fig:exp_aligned}(d), symmetric); a process that must be
driven by hydrodynamic interactions between the two bubbles, which lead to
an increase in the relative speed of the trailing bubble.

The hydrodynamic interactions arise from the changes in the bulk pressure field, which in the absence of the bubbles would decrease linearly along $x$. 
Single bubbles always propagate faster than the surrounding oil, as
seen in \S \ref{sec:single_bubble_systems}, with accompanying local
increases in pressure gradient. Hence, the fluid pressure at the front
and rear of a bubble propagating in the channel is respectively higher
and lower than the background pressure. The pressure
perturbation decays with distance from the bubble, but increases with
bubble volume because larger bubbles are faster. Consequently, when the smaller bubble is placed behind the larger, the net result of the perturbations due to both bubbles is that the trailing bubble experiences a lower local pressure near its tip, but a higher local pressure gradient causing the bubble to extend and narrow, see $t = 5.8$ in figure \ref{fig:exp_aligned}(b) and $t=8.3$ in panel \ref{fig:exp_aligned}(d). The resulting changes in bubble shape cause an increase in speed of the trailing bubble and eventually it catches the bubble in front. The trailing bubble's speed continues to increase as the bubbles approach because the local pressure gradient increases, which further modifies the bubble shape. The interaction just described is generic and has been observed in two-bubble interactions in other confined systems \citep{maxworthybubble1986,madecthesis2021}. The decay of the perturbations with distance means that if the bubbles are far enough apart the trailing bubble's speed does not increase sufficiently to allow it to catch the leading bubble.

 The transition between the separation and aggregation outcomes observed in figures~\ref{fig:exp_aligned}(a--d) was investigated by performing successive experiments with a variety of initial bubble distances $D(t=0)$. Results are shown in figure~\ref{fig:exp_aligned}(e), which presents the time evolution of the distance $D(t)$ between two initially asymmetric bubbles similar to that of figures~\ref{fig:exp_aligned}(a,b) for a variety of different initial $D(t=0)$. A transition between the two possible outcomes (aggregation in blue and separation in red) appears to occur for a threshold value of $D(t=0)$. All experiments, regardless of outcome, feature an initial decrease of $D(t)$ for $0<t<1.4$ which is associated with the rapid change in bubble shape following flow initiation; see e.g. figures~\ref{fig:exp_aligned}(a,b) where bubbles are more slender at $t=1.4$ than at $t=0$. This is followed by a monotonic increase in the case of separation or a steepening decrease in the case of aggregation. The neighbouring red and blue curves which bound the range of initial separations where the transition occurs feature an approximately flat region after their initial decrease, indicating that bubbles initially travel with approximately constant separation. This suggests the existence of an unstable two-bubble steady mode of propagation where the two bubbles would neither separate or aggregate but always remain at the same critical distance $D_{\mathrm{c}}$ from one another. We estimate $D_{\mathrm{c}}$ to be the average between the values of $D(t)$ for the (blue and red) curves adjacent to the threshold following initial decrease, i.e. $D(t=1.4)$ in figure~\ref{fig:exp_aligned}(e). This unstable state is a so-called edge state that marks the boundary between bubble separation and bubble aggregation.

\begin{figure}
	\centering
	\includegraphics[scale=0.6]{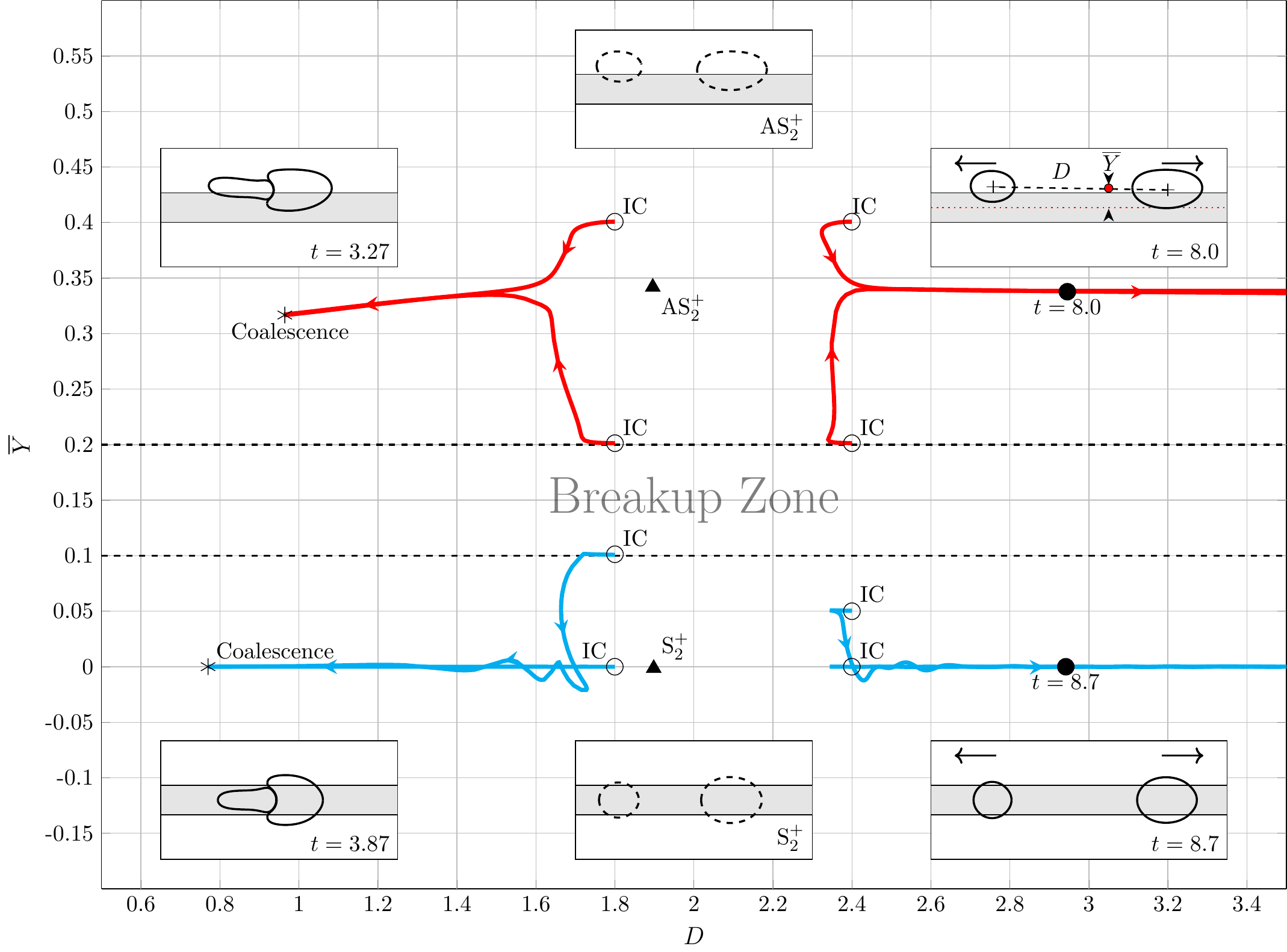}
	\caption{\small{Time-dependent calculations when $Q = 0.04>Q_{\mathrm{s}}$ so that larger bubbles are faster, $V_{\mathrm{r}} = 2/3, V_{\mathrm{total}} = \pi 0.54^2$. The lines are trajectories in the projected phase plane $(D,\overline{Y})$. Hollow circles indicate initial conditions (IC) and the arrows on the lines indicate increasing values of time, $t$. The triangles indicate the unstable steady states, shown as dashed bubble contours in the insets. Insets with solid line contours indicate the shapes of the bubbles at the stated time and correspond to solid markers on the trajectories. The dashed lines indicate a region where time-trajectories feature the breakup of at least one of the two bubbles.}}
	\label{fig:num_aligned_time_sim}
\end{figure}

 The evolution of two aligned bubbles in simulations of the theoretical model is very similar to that in the experiments. In figure~\ref{fig:num_aligned_time_sim}, time-simulations calculated for bubbles of volume ratio 2:1, propagating at flow rate $Q = 0.04$ from different aligned initial conditions are presented as trajectories in a projection of the phase space plotting the bubble separation, $D$, against the offset of the centre of mass $\overline{Y}$. Initial conditions with various initial global offsets $\overline{Y}(t=0)$ and separation distances $D(t=0)$ are denoted by hollow markers labelled `IC' and lead to either aggregation and then coalescence or separation of the two bubbles depending on the value of $D(t=0)$, as shown in the inset snapshots of the final outcomes with solid-line bubble contours.
Initial conditions with a global offset $\overline{Y}$ less than about $0.1$ ultimately lead to one or two symmetric bubbles $(\overline{y}_1 = \overline{y}_2 = 0)$ (blue curves) while initial conditions with a global offset $\overline{Y}$ larger than than about $0.2$ ultimately lead to one or two asymmetric bubbles (red curves). 
Moreover, as suggested by the experimental results, we find that there are unstable steadily propagating states in the model that divide the different dynamical outcomes.
The unstable steady states corresponding to the symmetric and asymmetric configurations are labelled $\mbox{S}_{2}^+$ and $\mbox{AS}_{2}^+$, respectively, and were calculated using the method detailed in appendix~\S\ref{sec:app_interval_bisection}.

\begin{figure}
	\centering
	\includegraphics[scale=0.6]{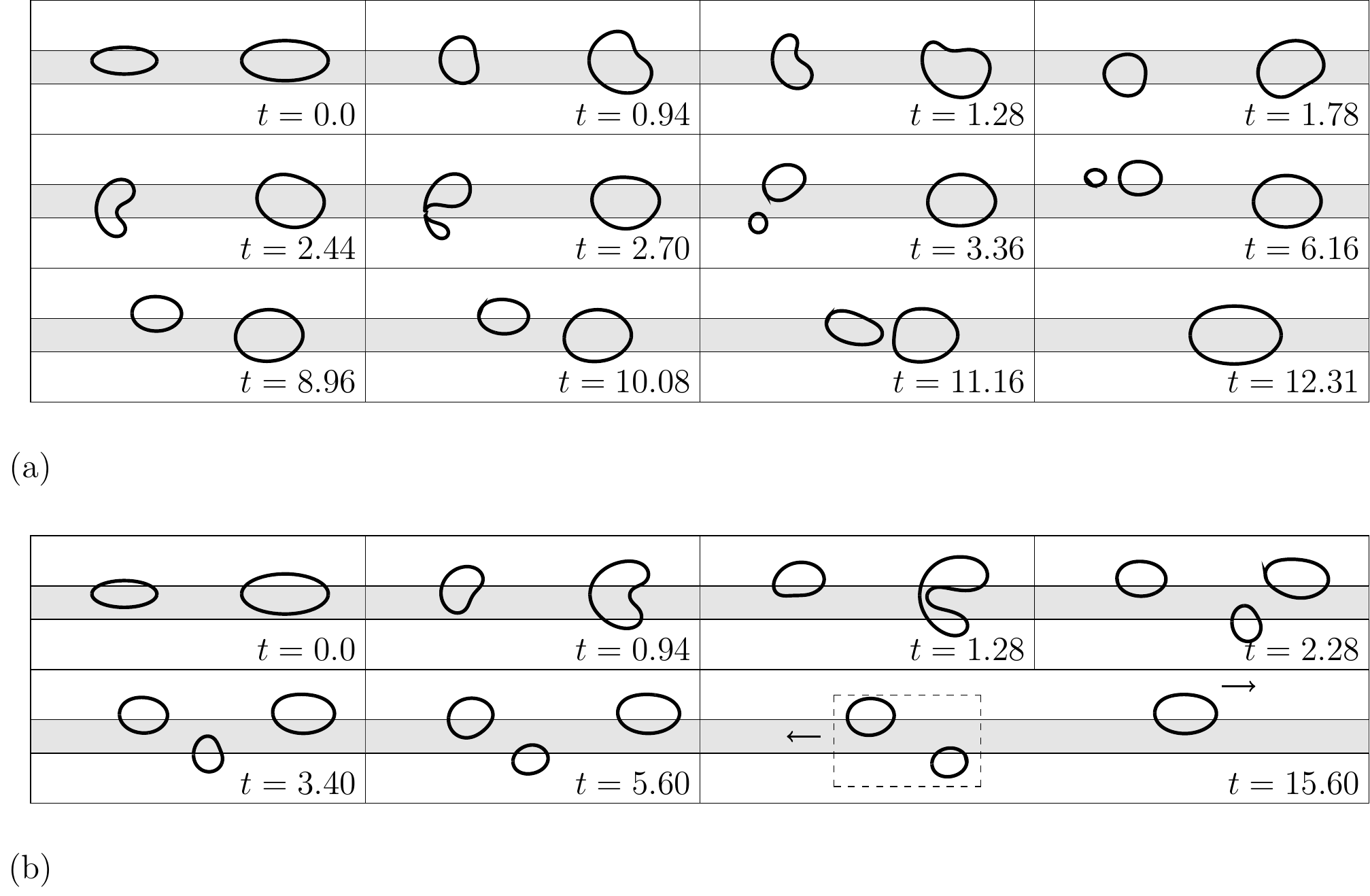}
	\caption{\small{Numerical time snapshots of the evolution of the system at times indicated in each panel, starting from initial conditions in the breakup zone in figure~\ref{fig:num_aligned_time_sim} at $Q=0.04$. Initial conditions are (a) $D=2.4,\overline{y}_1 = \overline{y}_2 = 0.10$ and (b) $D=2.4,\overline{y}_1 = \overline{y}_2 = 0.13$. In (b), the two trailing bubbles ultimately propagate at the same velocity, as indicated by the dashed box.}}
	\label{fig:3_bubble_zone}
\end{figure}

 For intermediate initial conditions, $0.1 \le \overline{Y}(t=0) \le 0.2$, we often observe bubble break up leading to three bubbles, as either a transient part of the evolution or a permanent outcome. Figure~\ref{fig:3_bubble_zone} shows two examples with initial global offsets $\overline{Y}(t=0) = 0.10$ and $0.13$ for the same initial separation distance $D(t=0) = 2.4$ in panels (a) and (b) respectively. In panel (a), the bubbles oscillate until the smaller trailing bubble breaks up before finally coalescing to form a single bubble which later coalesces with the leading bubble, ultimately generating a single steady symmetric bubble. The initial oscillations are reminiscent of the unstable periodic orbit identified in \cite{keeler2019invariant} for single bubbles. In panel (b), the leading bubble is initially more asymmetric and breaks up as it `hesitates' between an off-rail and on-rail configuration. However, here the bubbles do not coalesce and the final outcome is a three-bubble system with the larger leading bubble propagating faster than the two trailing bubbles which ultimately propagate steadily at the same speed on opposite sides of the rail. These two examples leading to two radically different final outcomes illustrate the sensitivity of the system to the initial bubble offset. The second example also opens up the possibility of a stable two-bubble state featuring offset bubbles, which will be explored in \S \ref{sec:offset}.

\begin{figure}
	\centering
	\includegraphics[scale=0.4]{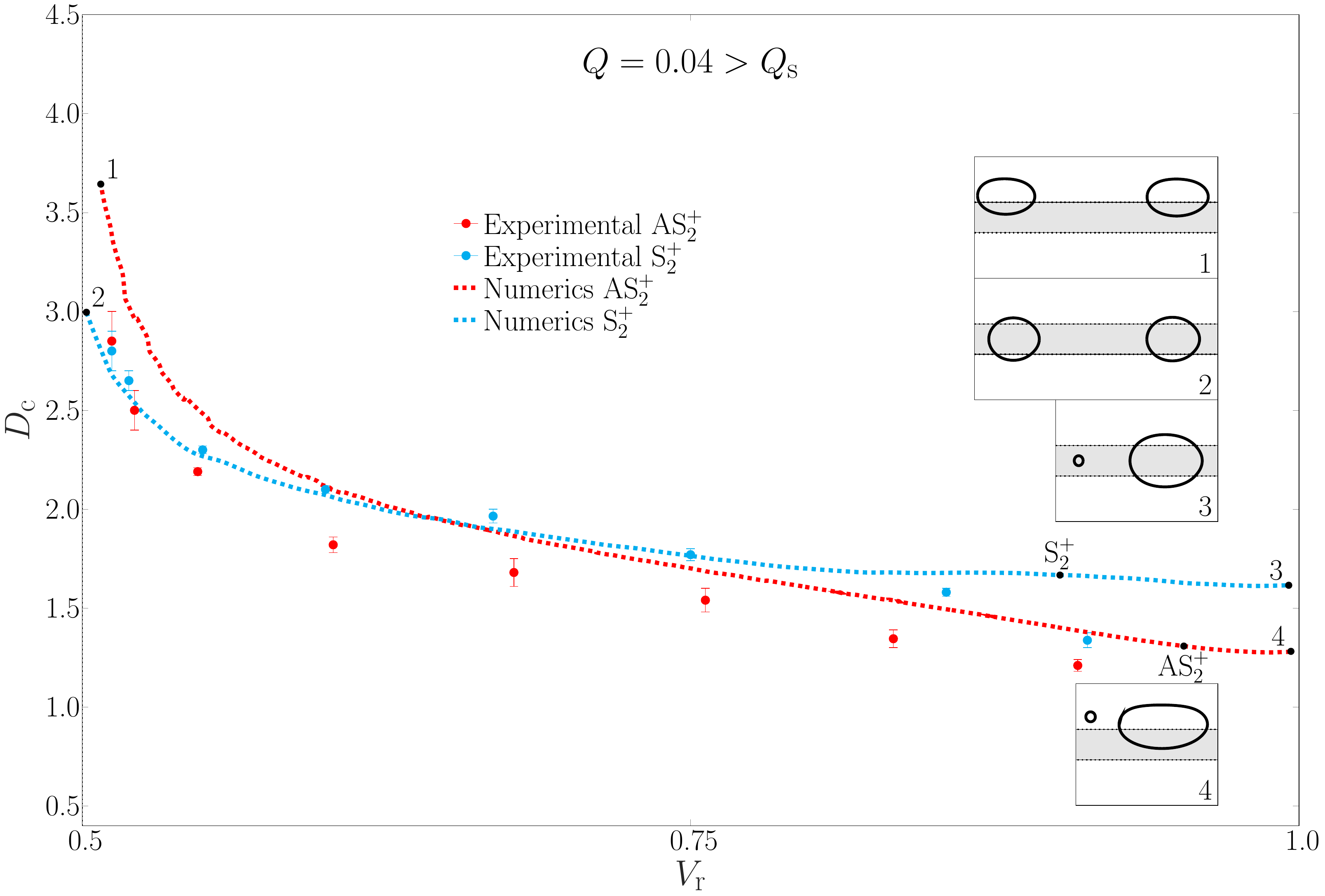}
	\caption{\small{Solution space as projected in the $(V_{\mathrm{r}},D_{\mathrm{c}})$ plane for the asymmetric $\mbox{AS}_{2}^+$ (red) and symmetric $\mbox{S}_{2}^+$ (blue) aligned edge states at fixed flow rate $Q=0.04$ and for a fixed total bubble volume $V_{\mathrm{total}}=0.54^2\pi$. Dashed lines correspond to numerical results while data points with error bars correspond to experimental values for which the bubble sizes as quantified by their area $A_i$ instead of their volume $V_i$ (see equation \eqref{eq:Vi}). Four snapshots of the numerical bubble shapes are shown for values of $V_{\mathrm{r}}$ indicated by black circular markers on the numerical lines labelled by digits from 1 to 4.}}
	\label{fig:num_aligned_Vr}
\end{figure}

Motivated by the fact that, as explored in \cite{gaillard2020life}, a single bubble can break up into bubbles of arbitrary volume ratio, we now use parameter continuation to determine the effect of the volume ratio $V_{\mathrm{r}}$ on two-bubble steadily propagating solutions. Figure~\ref{fig:num_aligned_Vr} shows the bubble separation distance $D_{\mathrm{c}}$ associated with the $\mbox{AS}_{2}^+$ (red) and $\mbox{S}_{2}^+$ (blue) edge states against the volume ratio $V_{\mathrm{r}}$ for a fixed total bubble volume and a fixed flow rate $Q=0.04$. The circular markers with error bars indicate experimental results and which are in reasonable agreement with numerical results. The agreement is generally within the experimental error for the symmetric states at smaller volume ratios, but the theoretical results consistently over-predict the separation distance for the asymmetric states, suggesting that the hydrodynamic interactions between two bubbles located near one edge of the rail are weaker in reality than in the model. The inset snapshots show the bubble configuration of the edge states at values of $V_{\mathrm{r}}$ indicated by numbered markers on the solution branches. For both edge states, the separation distance $D_{\mathrm{c}}$ decreases with increasing volume ratio and appears to converge to a finite value as $V_{\mathrm{r}}\to 1$ (i.e. $V_1/V_2 \to \infty$), see insets 3 and 4, while increasing sharply as $V_{\mathrm{r}}\to 1/2$ (i.e. $V_1/V_2 \to 1$), see insets 1 and 2. Linear stability results indicate that both branches are unstable with a single positive eigenvalue and that the least unstable eigenvalue approaches the imaginary axis as $V_{\mathrm{r}}\to 1/2$, indicating that the state with two equal bubbles is neutrally stable, which is consistent with the results of \cite{pumir1988}, and as expected because identical bubbles will travel at the same speed, assuming negligible hydrodynamic interactions. We note, however, that it is of course impossible in practice to have two bubbles of the exact same size in the experiments.

\begin{figure}
	\centering
	\includegraphics[scale=0.4]{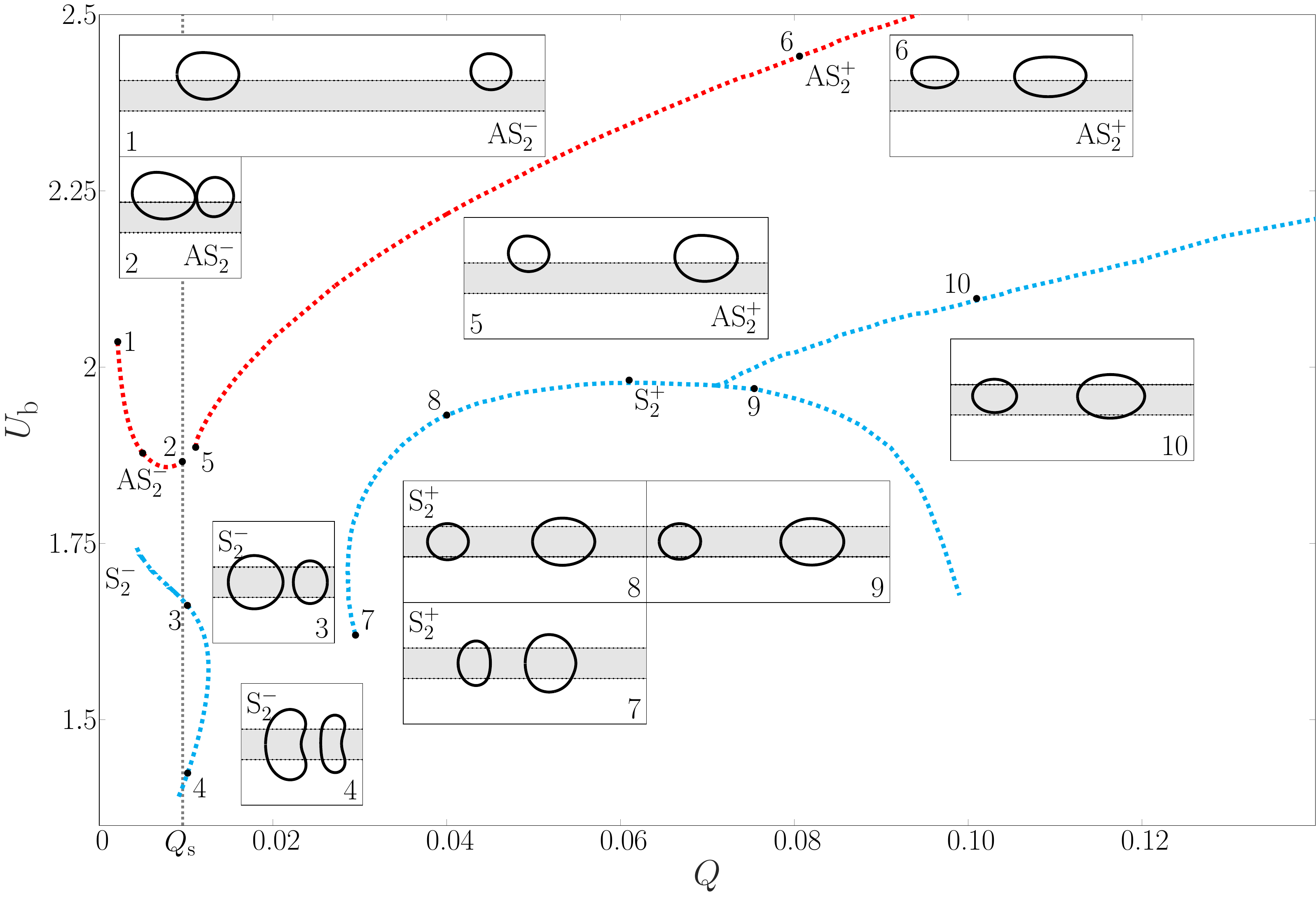}
	\caption{\small{Solution space for aligned bubbles, as projected in the $(Q,U_{\mathrm{b}})$ plane for a fixed bubble volume $V_{\mathrm{total}}=0.54^2\pi$ and bubbles of a volume ratio 2:1. The circular markers indicate solutions that are shown in the inset panels. Each inset has its own numerical label. Branches are dashed as the solutions are unstable and the dotted vertical line marks the position of $Q_{\mathrm{s}}\approx 0.0096$.}}
	\label{fig:num_aligned_solution_space}
\end{figure}

Figure~\ref{fig:num_aligned_solution_space} shows a bifurcation diagram of the different aligned two-bubbles states calculated through parameter continuation, where the velocity $U_{\mathrm{b}}$ of each state is plotted against $Q$ for a constant volume ratio $2:1$. Each solution branch is illustrated by at least one snapshot corresponding to a given value of $Q$ indicated by a circular marker on the branch. When $Q$ is larger than the transition flow rate $Q_{\mathrm{s}}$ discussed in \S\ref{sec:single_bubble_systems}, there are two branches discussed in figure~\ref{fig:num_aligned_time_sim} featuring symmetric ($\mbox{S}_2^+$) and asymmetric ($\mbox{AS}_2^+$) bubbles. In this case the symmetric $\mbox{S}_2^+$ branch only exists after a finite value of $Q\approx 0.03$ and experiences a pitchfork bifurcation, after which the two bubbles are slightly asymmetric, see inset labelled 10. We note that the asymmetric $\mbox{AS}_2^+$ branch persists for all values of $Q>Q_{\mathrm{s}}$ calculated but, as $Q$ approaches $Q_{\mathrm{s}}$ from above, the branch terminates as the bubbles become increasingly further apart and the limits of the computational domain are reached. For completeness, we also include unstable symmetric ($\mbox{S}_2^-$) and asymmetric ($\mbox{AS}_2^-$) solutions calculated for $Q<Q_{\mathrm{s}}$ where the leading bubble is now smaller than the trailing one. This is because the model predicts that for single steady bubbles, smaller bubbles propagate faster. However, as discussed in \S\ref{sec:single_bubble_systems}, numerical results for $Q<Q_{\mathrm{s}}$ do not reflect the experiments since there is no such transition flow rate in the experiments. The similarities between the two-bubble and single-bubble bifurcation diagrams presented in  figures~\ref{fig:num_aligned_solution_space} and \ref{fig:Qswitch} will be discussed in section~\ref{sec:comparing_1_2_bubble_solutions}.

\subsubsection{Offset bubbles}
\label{sec:offset}

\begin{figure}
	\centering
	\includegraphics[scale=0.3]{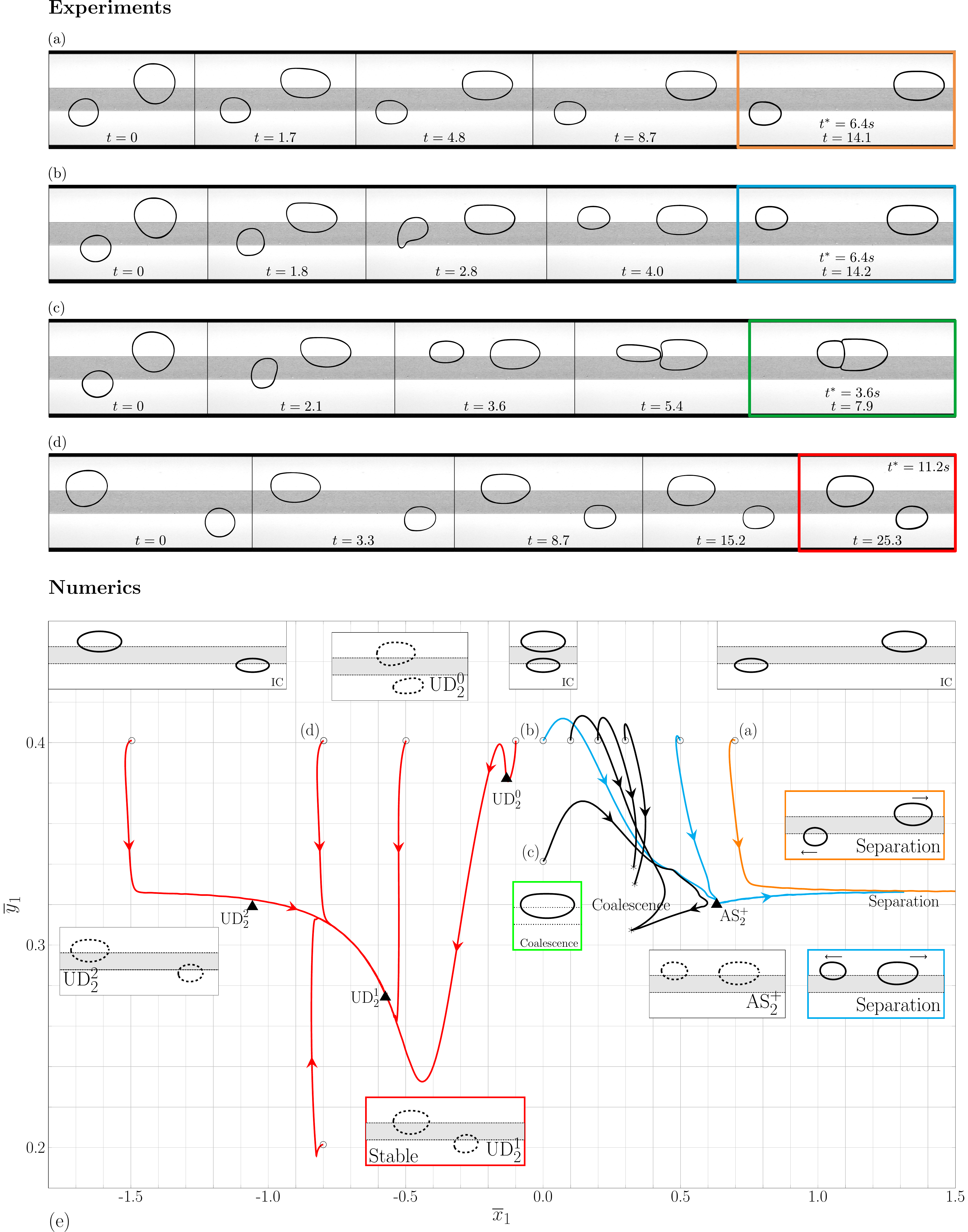}
	\caption{\small{(a-d): Experimental time snapshots of two bubbles propagating at flow rate $Q=0.04$ from initial offset configurations shown at t=0. The larger bubble is initially leading in (a-c), with decreasing separation distance from (a) to (c), and trailing in (d). Bubbles have a total area $A_{\mathrm{total}} = 0.54^2\pi$ and a 2:1 area ratio. The time labelling is the same as in figure~\ref{fig:exp_aligned}. (e): Corresponding numerical time-simulations presented as trajectories in a phase-plane projection using the coordinated $(\overline{x}_1, \overline{y}_1)$ of the larger bubble. The flow rate is $Q=0.04$ and bubbles have total volume $V_{\mathrm{total}} = 0.54^2\pi$ with a 2:1 volume ratio. Initial conditions are denoted by hollow circles, steady states by solid triangles and an star represents coalescence. Inset panels with label `IC' show three typical initial bubble configurations, inset panels with dashed bubble contours correspond to four steady states and inset panels with solid bubble contours and coloured outlines correspond to four different final outcomes for selected trajectories.}}        
	\label{fig:offset_time_evolution}
\end{figure}

We now consider offset bubble-pair configurations in which the two bubbles are initially positioned on opposite sides of the rail in an `Up-Down' (later denoted UD) configuration. The experimental protocol to prepare these configurations is outlined in appendix~\ref{sec:app_exp_protocol_offset}.

 In figures~\ref{fig:offset_time_evolution}(a--d), we present experimental time-sequences for two bubbles of area ratio 2:1 propagating at flow rate $Q = 0.04$ from different initial conditions. As in the case of aligned bubbles, if the larger bubble leads and the bubbles are initially well separated, figure \ref{fig:offset_time_evolution}(a), there is no significant hydrodynamic interaction and the bubbles separate indefinitely, remaining on their respective sides of the rail. As the distance between the bubbles decreases, then the hydrodynamic interaction between the bubbles is such that the trailing bubble migrates across the rail as it responds to the locally increased pressure gradient introduced by the leading bubble, see $t=4.0$ and $t=3.6$ in figures \ref{fig:offset_time_evolution}(b) and \ref{fig:offset_time_evolution}(c) respectively. Once the bubbles are on the same side of the rail, the system is in the aligned configuration, see \S\ref{sec:aligned}. The two bubbles separate indefinitely in figure \ref{fig:offset_time_evolution}(b) and aggregate in figure \ref{fig:offset_time_evolution}(c) owing to the different 
values of $D$ after bubble migration. 

In figure~\ref{fig:offset_time_evolution}(d), we consider the reverse initial configuration where the larger bubble is initially trailing. We observe that at first the trailing bubble propagates faster, as it would in the absence of hydrodynamic interaction, and hence the distance $D$ between the two bubbles decreases with time. As the bubbles approach, the trailing bubble starts to migrate over the rail, which causes it to slow down owing to the reduced viscous dissipation resulting from a smaller volume of fluid being displaced. In fact, the leading bubble also migrates further over the rail and so both bubbles slow down before ultimately reaching a steadily propagating state with a constant separation distance $D_{\mathrm{c}}$, see figure~\ref{fig:offset_time_evolution}(d) at $t=25.3$. This suggests the existence of a {stable} steady state of the two-bubble system, which is confirmed by numerical simulations. 

Corresponding numerical time-simulations are shown in figure~\ref{fig:offset_time_evolution}(e) and presented as trajectories in a $(\overline{x}_1,\overline{y}_1)$ projection of the phase space, where $(\overline{x}_1,\overline{y}_1)$ are the coordinates of the centroid of the largest bubble. We recall that the $x$-coordinate of the centre of mass of the system is constrained to be zero throughout calculations. The behaviour is similar to that observed in the experiments: in the case of an initially larger leading bubble ($\overline{x}_1>0$), the final outcome of the system switches from offset separation (orange trajectories) to aligned separation (blue trajectories) and finally to bubble coalescence (black trajectories) as the initial distance between the two bubbles is decreased. Figure~\ref{fig:offset_time_evolution}(e) shows that when the smaller trailing bubble crosses the rail, the $\mbox{AS}_2^+$ unstable steady state discussed in \S\ref{sec:aligned} acts as an edge state delineating aligned separation from coalescence outcomes, as blue and black trajectories approaching it from different sides are deflected in different directions.  In the case of an initially larger trailing bubble ($\overline{x}_1<0$), all (red) trajectories in figure~\ref{fig:offset_time_evolution}(e) converge towards a stable steadily propagating state labelled $\mbox{UD}_2^1$, irrespective of the initial distance between the bubbles and consistent with experimental observations.

We identify two further unstable steadily propagating states in figure \ref{fig:offset_time_evolution}(e), labelled $\mbox{UD}_2^0$ and $\mbox{UD}_{2}^2$. The red trajectory starting from the initial condition furthest to the right ($\overline{x}_1(t=0) = -0.1$) is first attracted towards the weakly unstable steady state $\mbox{UD}_2^0$. A transient bubble configuration extracted from such a time-simulation is used as an initial guess for calculating the $\mbox{UD}_2^0$ state. This state plays a role in transient dynamics involving two bubbles that are almost on top of each other, which occurs for example after breakup of a single bubble like in figure~\ref{fig:single_bubble_breakup} and was previously identified in \cite{gaillard2020life}, along with another unstable state labelled `Barrier' state that is not discussed in the present paper. In contrast, the state $\mbox{UD}_{2}^{2}$ appears to have no influence on the dynamics.

\begin{figure}
	\centering
	\includegraphics[scale=0.3]{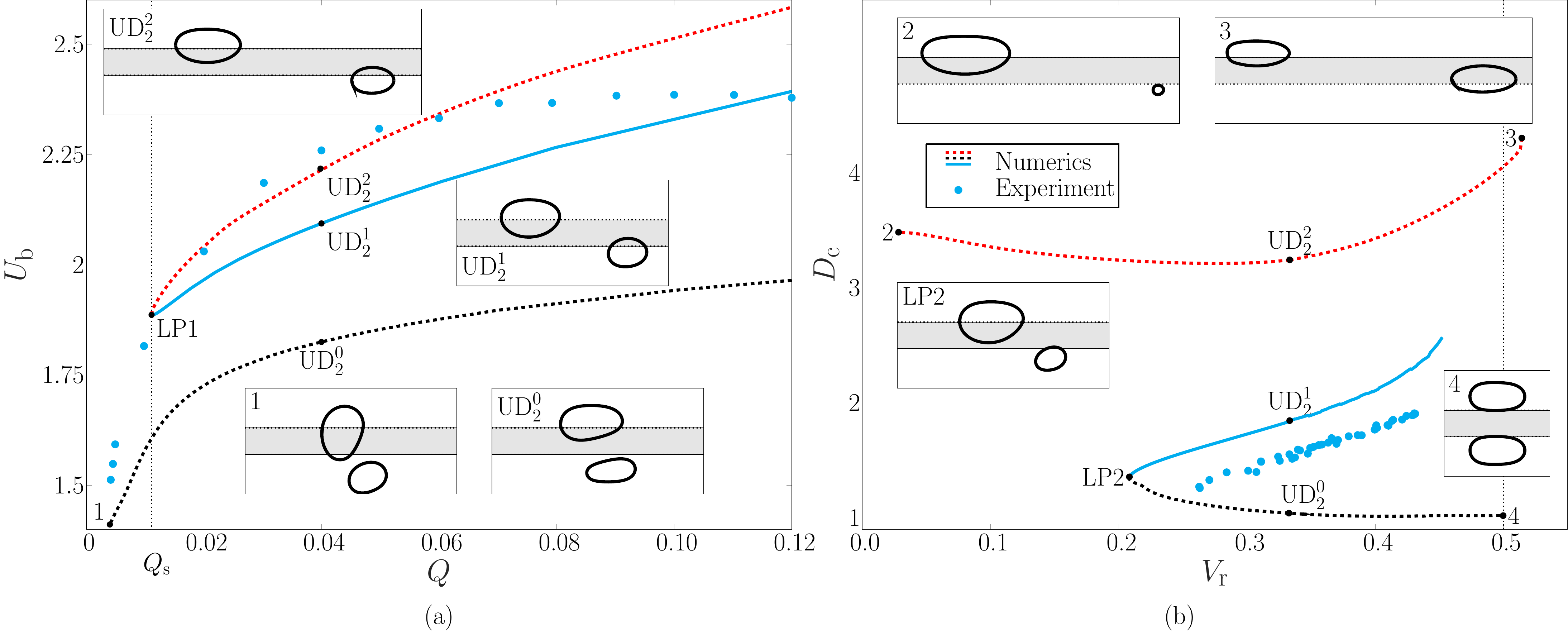}
	\caption{\small{Solution space for the offset steady states. Solid/dashed lines indicate stable/unstable solutions. The black circular markers indicate solutions that are shown in the inset panels whilst solid blue circular markers correspond to experimental results for the stable $\mbox{UD}_2^1$ state for bubbles of total area and area ratio equivalent to numerical bubbles volumes. (a) The solution space in the $(Q,U_{\mathrm{b}})$ plane when $V_{\mathrm{total}} = 0.54\pi^2$, $V_{\mathrm{r}} = 1/3$. The vertical dotted line indicates the value of $Q_{\mathrm{s}}$. (b) The solution space in the $(V_{\mathrm{r}},D_{\mathrm{c}})$ parameter plane for $Q=0.04$ and the same total volume. The vertical dotted line indicates when the two bubbles are the same size, i.e $V_{\mathrm{r}}=1/2$.}}
	\label{fig:offset_solution_space}
\end{figure}

\begin{figure}
	\centering
	\includegraphics[scale=0.37]{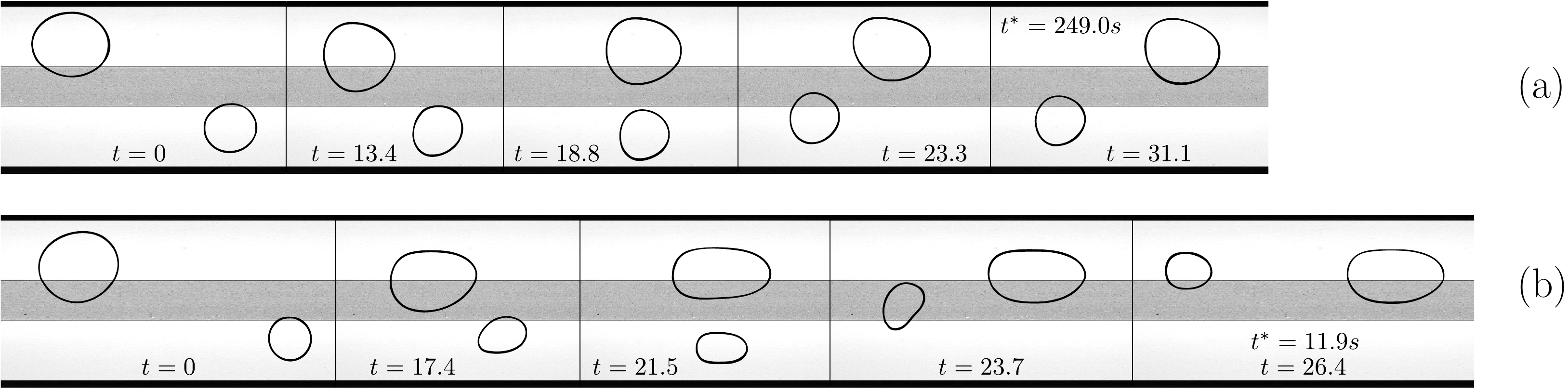}
	\caption{\small{Time sequence snapshots of experiments where $\mbox{UD}_2^1$ is not reached. (a) $A_{\mathrm{r}} \approx 1/3$ (bubble ratio 2:1) and $Q=0.002 < Q_{\mathrm{LP}1}$ and (b) $A_{\mathrm{r}} = 0.247$ and $Q=0.04$.}}
	\label{fig:exp_offset_limit_points}
\end{figure}

 We now use parameter continuation to examine the behaviour of the three offset two-bubble steady states $\mbox{UD}_2^0$, $\mbox{UD}_2^1$ and $\mbox{UD}_2^2$ as we vary the flow rate $Q$ and the bubble volume ratio $V_{\mathrm{r}}$. The two associated bifurcation diagrams are presented in figure~\ref{fig:offset_solution_space} where in panel (a) we present the velocity $U_{\mathrm{b}}$ associated with each state against $Q$ for a constant volume ratio $V_{\mathrm{r}} = 2/3$ and where in panel (b) we present the bubble separation distance $D_{\mathrm{c}}$ against $V_{\mathrm{r}}$ for a constant flow rate $Q=0.04$, similar to figures~\ref{fig:num_aligned_solution_space} and \ref{fig:num_aligned_Vr} respectively for the aligned bubble states. Solid and dashed lines indicate stable and unstable solutions respectively. Each branch is illustrated by at least one inset snapshot corresponding to a flow rate indicated by a circular marker on the branch. Experimental data points corresponding to measurements on the $\mbox{UD}_2^1$ stable state are also shown.  

 Figure~\ref{fig:offset_solution_space}(a) shows that the stable $\mbox{UD}_2^1$ and unstable $\mbox{UD}_2^2$ solutions are connected through a limit point denoted LP1.
 Unlike $\mbox{UD}_{2}^{0}$, the $\mbox{UD}_2^2$ unstable state appears to have no influence on the transient dynamics of the system according to figure~\ref{fig:offset_time_evolution}(e). The limit point LP1 occurs at a flow rate that coincides with the transition flow rate $Q_{\mathrm{s}} \approx 0.0096$, discussed in figure~\ref{fig:Qswitch}, where the mathematical model predicts that the smaller bubble propagates faster than the larger one in their respective AS1 single-bubble mode of propagation. This is consistent with the fact that a smaller leading bubble would then propagate faster in numerical time-simulations for $Q<Q_{\mathrm{s}}$ so that the two bubbles would separate without reaching a steady state. Surprisingly, although we know from \S\ref{sec:single_bubble_systems} that there is no evidence for $Q_{\mathrm{s}}$ in the experiments, a threshold, consistent with a limit point, of the $\mbox{UD}_2^1$ state is also found experimentally at a critical flow rate $Q_{\mathrm{LP}1} = 0.0040 \pm 0.0002$ for a bubble area ratio $A_r \approx 1/3$. A representative time-evolution observed experimentally for $Q<Q_{LP1}$ is shown in figure~\ref{fig:exp_offset_limit_points}(a). In this case, the migration of the trailing bubble over the rail induced by the bubble interaction does not cause a sufficient speed reduction to reach a steadily propagating two-bubble state. Instead, the larger bubble passes over the smaller bubble so that both bubbles ultimately propagate steadily and separate indefinitely. This low flow rate scenario is not captured using our model. We also note that the $\mbox{UD}_2^0$ branch in figure~\ref{fig:offset_solution_space}(a) is distinct from the two others and could not be calculated numerically below a flow rate indicated by the label `1' at which the smaller bubble touches the side wall of the channel, see associated inset snapshot.

 Figure~\ref{fig:offset_solution_space}(b) shows that the stable $\mbox{UD}_2^1$ and unstable $\mbox{UD}_2^0$ solutions, which were disconnected in the $(Q,U_{\mathrm{b}})$ projection of figure~\ref{fig:offset_solution_space}(a), are in fact connected through a limit point denoted LP2 under variations in volume ratio. This means that there is a critical volume ratio below which the stable $\mbox{UD}_2^1$ state does not exist, which we find to be $V_{\mathrm{r}} \approx 0.21$ at $Q=0.04$. The existence of such a critical volume ratio is supported by our experiments, in which no stable state is found for an area ratio $A_r < 0.256 \pm 0.007$ at the same flow rate. A representative experimental time-evolution at $A_{r} = 0.247$ is shown in figure~\ref{fig:exp_offset_limit_points}(b) and equivalent numerical simulations are qualitatively similar. The overall dynamics are essentially the same as those at low $Q$ below LP1 in the experiments: for values of the volume ratio below LP2 the interaction between the bubbles does not reduce the speed of the trailing bubble sufficiently to establish a steadily propagating two-bubble state. 
The only qualitative difference to the time evolution shown in  figure~\ref{fig:exp_offset_limit_points}(a) is that
the smaller bubble migrates over the rail once the larger bubble has moved ahead.

Figure~\ref{fig:offset_solution_space}(b) also shows that, like in figure~\ref{fig:num_aligned_Vr} for aligned bubbles, the distance $D_{\mathrm{c}}$ between the bubbles increases when approaching the limit of two bubbles of equal sizes ($V_{\mathrm{r}} = 1/2$) for the $\mbox{UD}_2^1$ and $\mbox{UD}_2^2$ branches. The chosen fixed length of the computational domain means that the $\mbox{UD}_2^1$ branch solution could not be calculated close to that limit. By contrast, the $\mbox{UD}_2^0$ solution could be calculated up to the $V_{\mathrm{r}}=1/2$ limit where it features two identical bubbles on either side of the rail propagating at the same $x$-position with opposite offsets $y_2 = - y_1$.

\subsection{Comparison of single- and two-bubble solution structures}
\label{sec:comparing_1_2_bubble_solutions}

\begin{figure}
	\centering
\includegraphics[scale=0.3]{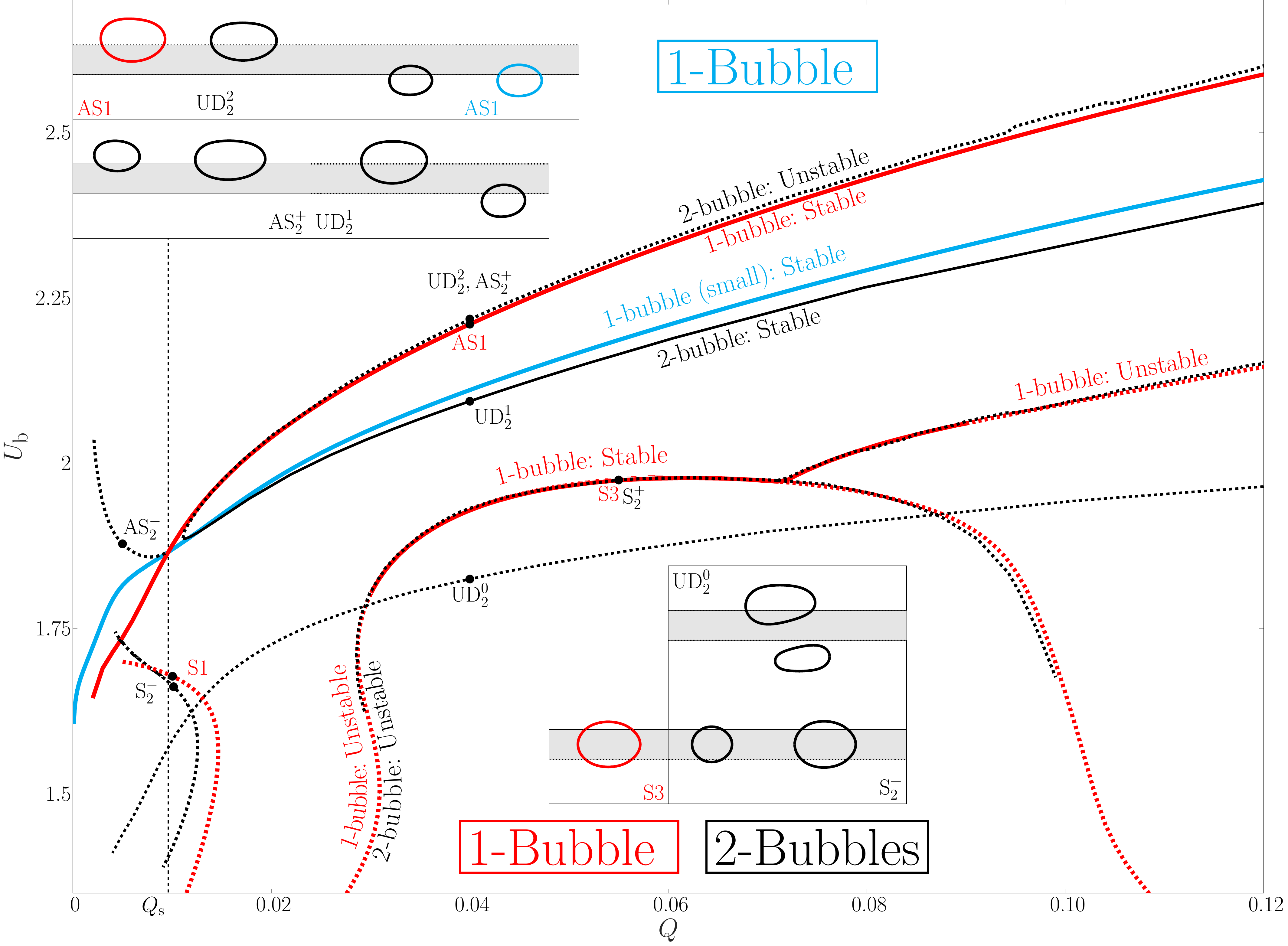}
\caption{\small{Comparison of the single-bubble and two-bubble solution spaces plotted in the $(Q,U_{\mathrm{b}})$ plane. The single bubble space is denoted by coloured lines and the two-bubble system by black lines, with solid/dashed lines indicating stable/unstable branches. The AS1 solution branch for the single-bubble system is shown for the larger (red) and smaller (blue) bubble. Solid markers indicate specific solutions on the branch shown in the inset panels. The vertical dotted line indicates $Q_{\mathrm{s}}$. The volume of the single-bubble is identical to the larger of the bubbles in the two-bubble system.}}
	\label{fig:comparison}
\end{figure}

There is a striking similarity between the solution structure for single bubbles presented in figure~\ref{fig:Qswitch} and discussed in our previous papers \citep{keeler2019invariant,gaillard2020life}, and the solution structure for two bubbles presented in figures~\ref{fig:num_aligned_time_sim} and \ref{fig:offset_solution_space}. For ease of comparison, figure~\ref{fig:comparison} shows a direct comparison between the solutions in the $(Q,U_{\mathrm{b}})$ plane for a single bubble (coloured lines) and a two-bubble system (black lines). The total bubble volume in the two-bubble system is $V_{\mathrm{total}} = \pi0.54^2$ and the bubble volumes have a ratio 2:1. The bubble volume in the single-bubble system is the same as the volume of the larger bubble of the two-bubble system, i.e. $V=2/3 V_{\mathrm{total}}$. A selection of the branches are illustrated by snapshots at given flow rates indicated by a circular marker.

For $Q>Q_{\mathrm{s}}$ the two-bubble $\mbox{AS}_2^{+}$ and $\mbox{UD}_2^2$ solution branches overlap and closely match the single-bubble asymmetric state, labelled $\mbox{AS}_1^{+}$. Furthermore the two-bubble symmetric $\mbox{S}_2^{+}$ solution branch is almost indistinguishable from the symmetric one-bubble state, labelled $\mbox{S}_1^{+}$. For $Q<Q_{\mathrm{s}}$ there is no single-bubble state corresponding to $\mbox{AS}_2^{-}$, but the symmetric two-bubble $\mbox{S}_2^{-}$ solution branch has a similar structure to the single-bubble $\mbox{S}_1^-$ branch, albeit without the excellent quantitative agreement found in $Q>Q_{\mathrm{s}}$ case. We examined variations in the single-bubble volume and found that the closest match between the single and two-bubble solution branches for $Q>Q_{\mathrm{s}}$ occurs when the single bubble volume is chosen to be equal to that of the leading bubble of the two-bubble system, as shown in figure~\ref{fig:comparison}, which indicates that the leading bubble (not necessarily the fastest) sets the speed of the bubble pair. This comparison is particularly evident when comparing two-bubble branches with a larger leading bubble (see the $\mbox{AS}_2^+$ and $\mbox{S}_2^+$ and their single-bubble counterparts) but, as can be seen from the figure, the stable two bubble $\mbox{UD}_2^1$ state is slightly slower than the corresponding AS1 state corresponding to leading (smaller) bubble. 
  
 \begin{figure}
   \centering
    \includegraphics[scale=1.2]{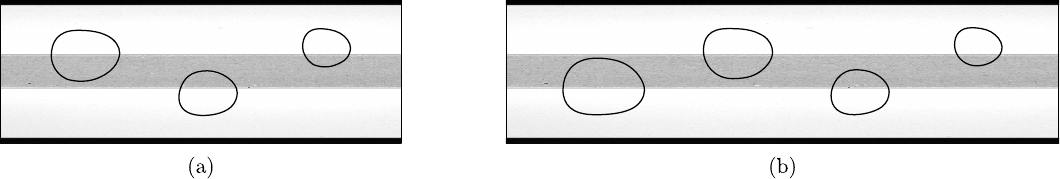}
	 \caption{\small{Experimental evidence of a multi-bubble stable steady states, for $Q = 0.011$. (a) A 3-bubble stable steady state (dimensionless volumes $0.33^2\pi, 0.40^2\pi, 0.47^2\pi$) (b) A 4-bubble stable steady state (dimensionless volumes $0.33^2\pi, 0.40^2\pi, 0.47^2\pi, 0.54^2\pi$).}}
   \label{fig:three_bubble_edge}
 \end{figure}

\section{Discussion}
\label{sec:discussion}
 In this paper we studied the propagation of two bubbles through a geometrically perturbed Hele-Shaw channel under constant flow rate: the simplest configuration that introduces bubble-bubble interactions in the system. The study is of fundamental interest in a variety of applications, but our initial motivation was to conduct a controlled investigation of the post break-up dynamics of the single bubbles that we studied in our previous paper \citep{gaillard2020life}.
When a single bubble breaks up, the relative positions and volumes of the resulting multiple bubbles are very sensitive to perturbations in the system and are extremely difficult to control. In this paper, we fixed the sum of the two-bubble volumes, but varied the relative sizes of each bubble to simulate different break-up configurations.

We find that the general behaviour of the two-bubble system falls into four different long-term outcomes: (i) indefinite separation of the two bubbles; (ii) aggregation and coalescence of the two bubbles; (iii) a steadily propagating two-bubble state and (iv) break up to form a larger number of bubbles with potential future break-up and aggregation/coalescence events. We demonstrate that, as in the case of the single bubble, the overall dynamics is orchestrated by both stable and unstable steadily-propagating one- and two-bubble states. The ranges of existence and stability of the states depend on the flow rate and relative sizes of the two bubbles. The general behaviour of the system is qualitatively described by a depth-averaged theory, provided that the flow rate is sufficiently large, $Q > Q_{\mathrm{s}}$.

The existence of steadily-propagating two-bubble states is striking
because they have not been in observed in other confined systems
\citep{maxworthybubble1986,madecthesis2021}, in which the multiple
bubbles will always separate or aggregate. There are large regions of
overlap between particular single- and two-bubble steadily propagating
states in the relationship between bubble speed and flow rate for a
fixed volume ratio between the two bubbles, see
figure~\ref{fig:comparison}. The overlap regions suggest that these
single- and two-bubble states are closely related. In particular, over
a wide range of flow rates, the unstable asymmetric two-bubble states
$\mbox{UD}_{2}^{2}$ and $\mbox{AS}_{2}^{+}$ have the same speed as the
stable asymmetric single bubble $\mbox{AS}_{1}$; and the unstable
symmetric two-bubble state $\mbox{S}_{3}$ has the same speed as
the stable symmetric single-bubble $\mbox{S}_{1}^{+}$. In these
comparisons, the single bubble always has the same volume as the
larger of the two bubbles in the two-bubble state, indicating that the
two-bubble state moves at the speed of the leading bubble and that the
leading bubble is not significantly affected by the interaction. In
other words, the leading bubble is driving the dynamics and the
trailing bubble is carried along with it. This is only possible
because the presence of the rail allows the trailing bubble to
experience different local geometric confinements depending on its
lateral position within the channel.

 Having established the existence of two-bubble states, we can extend
 the methods used in this paper to construct a variety of multiple
 bubble states and the number of possible states increases
 dramatically with the number of bubbles, in line with the number of
 permutations of increasing numbers of discrete objects.
 Experimental confirmation of the existence of what appear to be
 stable three- and four-bubble steadily propagating states is given in figure~\ref{fig:three_bubble_edge}.
 The existence of stable and unstable $n$-bubble steadily propagating
 states will have a potential influence on the dynamics of bubble
 trains in confined systems \citep{beatusphonos2006,beatusdroplets2012}
 in the presence of  imperfections in both geometry and bubble volume. The ranges of
 existence of the multiple-bubble solutions and their sensitivity to
 perturbations will be pursued in a future investigation.

 Finally, although there is not a direct equivalent of the
 $\mbox{UD}_2^{0}$ state in the single-bubble system, the
 $\mbox{UD}_{2}^{0}$ state is the only two-bubble state that persists if the
 height of the rail, $h$, is decreased to 0. All of the other
 steady states cease to exist because their separation distances
 increases as $h\to 0$, but the $\mbox{UD}_2^{0}$ state barely changes and exists exactly when $h=0$. The question of whether a stable two-bubble steady state exists when $h = 0$ is an open question but certainly if one does exist it is is unlikely to be related to the steady states constructed here. Indeed, numerical IVP calculations confirm that a two-bubble steady state does not exist in the same form as the $\mbox{UD}_2^1$ solution (smaller bubble head on opposite sides) and that the dynamics are incredibly slow, indicating neutral stability. This exploration of two-bubble steady states in the experiment will form part of a future investigation.


\appendix

\section{Experimental Protocols}
\label{sec:app_exp_protocols}

\subsection{Experimental Protocol: Aligned Bubbles}
\label{sec:app_exp_protocol_aligned}

Aligned pairs of bubbles were obtained by producing two bubbles sequentially with the trailing bubble generated once the leading bubble had propagated by a prescribed distance from the air port, which was set by the volume of oil injected in the intervening time. A dimensionless flow rate $Q_i$ was then imposed to propagate the pair of bubbles to a position a few centimetres downstream of the centring device before interrupting the flow for half a second. For $Q_i=0.029$, the two bubbles straddled the rail symmetrically about the channel centreline. However, asymmetric bubbles which overlapped the rail from only one side could also be obtained for sufficiently small values of the flow rate ($Q_i = 0.0056$) because of the absence of stable one-bubble symmetric modes of propagation, which meant that initially centred bubbles migrated off the rail. This migration was always to the same side of the rail because of unavoidable bias in the levelling of channel; see \citet{gaillard2020life} for details. 

\subsection{Experimental Protocol: Offset Bubbles}
\label{sec:app_exp_protocol_offset}

The simplest method to generate bubbles on opposite sides of the rail is to break up a single bubble into two parts, as illustrated in figure~\ref{fig:single_bubble_breakup}(b). This was achieved by initially propagating a single symmetric bubble downstream of the centring device at a flow rate $Q_i = 0.03$ before interrupting the flow for a controlled duration to allow the bubble to widen and initiate its sideways migration towards one of the deeper regions of the channel, see \citet{gaillard2020life} for a description of the bubble relaxation process at $Q = 0$. A flow rate of $Q = 0.05$ was then imposed which led the bubble to break into two parts of different sizes, overlapping the rail from opposite sides. The difference in bubble sizes after breakup was set by the $y$-offset of the single bubble upon imposing the flow. Depending on the direction of the oil flow used to break up the initial bubble (either from the inlet towards the outlet or vice-versa), different initial bubble-pair configurations could be prepared, with the larger bubble either leading (closer to the outlet) or trailing (closer to the inlet). The initial separation between the two bubbles was controlled by imposing a flow rate of $Q = 0.007$ in the same direction as the flow used to split the single bubble. This flow rate was chosen for two reasons. Firstly, at this flow rate, the larger leading bubble propagated faster than the smaller trailing bubble so that the distance between the two bubbles increased with time. Secondly, this flow rate was also sufficiently small to prevent the smaller bubble from crossing the rail because of its hydrodynamic interaction with the larger bubble. At lower flow rates, the relative strength of the capillary force acting on each bubble increased which in turn meant that they propagated more asymmetrically, ensuring that both bubbles remained on their respective sides of the rail. Finally, when the bubbles reached a prescribed separation, the flow was interrupted for half a second before beginning the two-bubble experiment.

\section{Numerical Methods}
\label{sec:app_numerical_methods}

\subsection{Finite-element method}
\label{sec:app_finite_elements}

The system of equations are solved using the finite-element method using the open-source \texttt{oomph-lib} library \citep{heil2006oomph}. An unstructured triangular mesh is fitted to the boundaries in the computational domain and the pressure unknowns in the fluid are interpolated using piecewise quadratic functions. The position of the bubble contours are unknown in the problem and a pseudo-solid node update procedure is used to facilitate this. The position of the bubble boundaries are found by introducing an additional unknown Lagrange multiplier field to the solid equations on the bubble which are determined by the dynamic boundary condition whilst the kinematic boundary condition is incorporated naturally in the weak form of the equations. The resulting sets of equations are solved using Newton's method. During time-dependent calculations a Backwards Euler method is used with a typical time-step of $\Delta t = 0.01$. Every five steps we adapt the underlying spatial mesh using a ZZ error estimator based on continuity of pressure gradients between elements \citep{zienkiewicz1992}. If the error is below a certain tolerance the mesh is unrefined (typically chosen as $\sim 10^{-6}$) and if above a certain tolerance the mesh is refined (typically chosen as $\sim 10^{-3}$). In both steady and time-dependent calculations, these tolerances are adjusted to check that the solution has converged. Typically there are $~5000$ elements in the mesh, the smallest is of order $~10^{-5}$ and the largest $~10^{-2}$ (based on the bubble volumes chosen in this paper).

\subsection{Topology Changes}
\label{sec:app_topology_changes}
Figure~\ref{fig:3_bubble_zone} shows the numerous topology changes the system of bubbles experience. To facilitate topology change (i.e. bubble coalescence and bubble breakup) in the numerical code we employ the procedure described in \cite{gaillard2020life}. After each timestep, we measure the minimum distance between each of the pairs of bubbles and also check if each individual bubble has self-intersected. If the minimum distance between two bubbles is lower than a pre-defined threshold (in these calculations we choose $10^{-2}$), the bubbles are merged into a single bubble. Alternatively, if self-intersection has been identified we split the bubble into two separate bubbles. In each case, the simulations continue after the topology change, with the number of volume constraints in the system deleted/added, as appropriate. For more details we refer the reader to \cite{gaillard2020life}.

\subsection{Interval-Bisection Algorithm}
\label{sec:app_interval_bisection}

The method for calculating the aligned states is now described in more detail. The algorithm is initiated by solving two IVPs, one in the case where the two bubbles are initially sufficiently far from one another to separate and one where they are sufficiently close to one another to ultimately coalesce, as illustrated in figure~\ref{fig:num_aligned_time_sim}. After this initial step, a new IVP is solved where the initial distance between the bubbles is
\begin{equation}
  D_{\mathrm{edge}}(t = 0) = (D_{\mathrm{s}} + D_{\mathrm{c}})/2,
  \label{bisect}
\end{equation}
where $D_{\mathrm{s}}$ and $D_{\mathrm{c}}$ are the values of the initial bubbles distance in the previous simulations that lead to separation and coalescence respectively. Once the final dynamical outcome is established, for this IC, either $D_{\mathrm{s}}$ or $D_{\mathrm{c}}$ is updated to the value of $D_{\mathrm{edge}}(t = 0)$, as appropriate, and a new value of $D_{\mathrm{edge}}(t = 0)$ is chosen from \eqref{bisect}.

This interval-bisection procedure is repeated so that the initial bubble distance $D_{\mathrm{edge}}$ converges to a value so that $D(t)\to D_{\mathrm{crit}}$ corresponding to the unstable steady state. In each simulation, the volume and the initial offset and shape (slender ellipse) of each bubble is kept constant while only varying the initial distance between the two bubbles. The final dynamical outcome is determined when either the minimum distance between the two bubble contours gets smaller than a cutoff value $D(t) < D_{\mathrm{min}}$, which is small enough to ensure that the bubbles will coalesce, or when the centroid-to-centroid distance $D(t)$ gets larger than a cutoff value $D(t) > D_{\mathrm{max}}$ which ensures that the bubbles will separate indefinitely. We use values of $D_{\mathrm{min}} = 0.01,D_{\mathrm{max}} = 3$ for the results of this paper.

The convergence criteria for the interval-bisection procedure is that the bubble remain within a small distance, $\varepsilon$, of each other for a large time, $T$, i.e.
\begin{equation}
	|D(t) - D_{\mathrm{edge}}(t=0)|<\varepsilon,\quad  \forall t<T_{\mathrm{end}}.
\end{equation}
Once this condition has been satisfied we solve the steady governing equations to get the unstable steady state. In the results presented here find that $\varepsilon = 0.1$ and $T=20$ is sufficient to ensure that we converged to a steady state.

\section{Acknowledgments}

We acknowledge funding from the Engineering and Physical Sciences Research Council through and grant numbers EP/P026044/1 EP/T021365.

\bibliographystyle{RS} 
\createbib{hele_shaw_paper}

\end{document}